\documentclass[twocolumn]{aastex62}
\usepackage{amsmath}     
\usepackage{wasysym}           
\usepackage{graphicx}
\usepackage{amssymb}
\usepackage{epstopdf}
\usepackage{mathrsfs}
\usepackage{anyfontsize}
\usepackage{natbib}
\usepackage{color}
\usepackage{lipsum}
\usepackage{diagbox}

\shorttitle{On the relativistic rate of tdes} 
\shortauthors{Coughlin \& Nixon}
\begin{document}
\title{On the impact of relativistic gravity on the rate of tidal disruption events}

\author[0000-0003-3765-6401]{Eric R.~Coughlin}
\affiliation{Department of Physics, Syracuse University, Syracuse, NY 13210, USA}

\author[0000-0002-2137-4146]{C.~J.~Nixon}
\affiliation{Department of Physics and Astronomy, University of Leicester, Leicester, LE1 7RH, UK}

\email{ecoughli@syr.edu}

\begin{abstract}
The tidal disruption of stars by supermassive black holes (SMBHs) probes relativistic gravity. In the coming decade, the number of observed tidal disruption events (TDEs) will grow by several orders of magnitude, allowing statistical inferences of the properties of the SMBH and stellar populations. Here we analyse the probability distribution functions of the pericentre distances of stars that encounter an SMBH in the Schwarzschild geometry, where the results are completely analytic, and the Kerr metric. From this analysis we calculate the number of observable TDEs, defined to be those that come within the tidal radius $r_{\rm t}$ but outside the direct capture radius (which is, in general, larger than the horizon radius). We find that relativistic effects result in a steep decline in the number of stars that have pericenter distances $r_{\rm p} \lesssim 10\,r_{\rm g}$, where $r_{\rm g} = GM/c^2$, and that for maximally spinning SMBHs the distribution function of $r_{\rm p}$ at such distances scales as $f_{\rm r_{\rm p}}\propto r_{\rm p}^{4/3}$, or in terms of $\beta \equiv r_{\rm t}/r_{\rm p}$ scales as $f_{\beta} \propto \beta^{-10/3}$. We find that spin has little effect on the TDE fraction until the very high-mass end, where instead of being identically zero the rate is small ($\lesssim 1\%$ of the expected rate in the absence of relativistic effects). Effectively independent of spin, if the progenitors of TDEs reflect the predominantly low-mass stellar population and thus have masses $\lesssim 1M_{\odot}$, we expect a substantial reduction in the rate of TDEs above $10^{7}M_{\odot}$. 
\end{abstract}

\keywords{Black hole physics (159) --- Event horizons (479) --- General Relativity (641) --- Kerr black holes (886) --- Relativistic mechanics (1391) --- Tidal disruption (1696)}

\section{Introduction}
At a rate of $\sim 10^{-4}-10^{-5}$ per year, unfortunate stars make their way to the centre of each galaxy, passing too close to the central supermassive black hole (SMBH) to survive the encounter. Upon reaching the tidal radius, where the SMBH's tidal field and the star's self-gravity become approximately equal, the star is disrupted and stretched into a stream of debris. Some of this debris can return to the SMBH to form an accretion flow that powers a luminous transient called a tidal disruption event (TDE). There have now been $\gtrsim 50$ TDEs observed to date \citep{gezari21}, and this number is expected to increase substantially in the next few years \citep[e.g.,][]{bricman20}. 

Theoretical predictions for the rate of TDEs, starting with \cite{frank76}, have been remarkably consistent for many years now, finding $\sim 10^{-4}-10^{-5}$ per year per galaxy \citep[e.g.,][]{magorrian99,wang04,merritt13,stone16,zhong22}. It is possible that some non-steady stellar systems can produced periods of enhanced rates for short durations \citep[e.g.][]{madigan18}. The observed rate of TDEs is starting to reflect a similar value \citep[e.g.,][see also the discussion in \citealt{gezari21}]{hung17,vanvelzen20}. So it seems that from a broad perspective there is reasonable agreement between the theoretical predictions and observed rates for the average number of TDEs per galaxy in the Universe.

The expected increase in the number of observed events in the coming years provides a hope that we will be able to discern astrophysical quantities from TDE statistics. Modelling TDE lightcurves has the potential to reveal (1) the properties of the SMBH (mass and spin), (2) the properties of the star that is being disrupted (mass, age, metallicity, spin, multipliticity)\footnote{And potentially whether the star hosts planets as the presence of planets will lead to (1) a phase-dependent energy shift in the stellar orbit near pericentre and (2) an increase in the fallback rate and variation in fallback composition when the planetary material enters the fray (as the density of planets is $\sim$ the density of stars it is likely that if the star is disrupted then so is the planet; although for planets orbiting the star at radii greater than the tidal radius it is also possible that they are ejected as hyper-velocity planets [or HVPs; \citealt{ginsburg12}] without being disrupted). However, the observable consequences of planets in stellar TDEs is expected to be quite small, so this would require an incredible level of precision in both modelling and observed data.}, and (3) the orbital properties of the star around the SMBH \citep[see the modelling efforts of, e.g.,][]{lodato09, guillochon13, gafton15, shiokawa15, bonnerot16, golightly19b, sacchi19, golightly19a,lawsmith19}. By combining detailed modelling of individual events with statistical inferences from a large number of TDEs, it will become possible to infer the distribution functions of these properties, which will significantly increase our understanding of several important astrophysical processes.

However, our ability to infer these properties relies on a detailed understanding of the physics underlying these events. In this paper we revisit, explore and revise calculations of the dependence of TDEs on the relativistic nature of the gravitational field generated by the SMBH \citep[e.g.,][]{beloborodov92,kesden12,will12,servin17}. Relativistic gravity has important differences from a Newtonian description and in some regions of parameter space these can lead to significant revision of the inferred TDE rates \citep[e.g.,][]{kesden12}. Relativistic gravity introduces the concept of a direct capture radius, inside of which the stellar debris is not able to recede away from the SMBH and subsequently form an accretion flow, but is instead captured and swallowed by the SMBH; we emphasize that this is distinct from, and generally larger than, the horizon radius. Relativistic gravity also generates a stronger tidal field experienced by the star at a given radius compared to the Newtonian estimate. This results in a larger fraction of stars entering the direct capture radius and thus a reduced number of TDEs\footnote{Presumably even stars that are directly captured by the SMBH are tidally disrupted prior to hitting the singularity, but for simplicity of notation we refer to stars that enter within the direct capture radius as ``direct capture events'' and those that enter within the tidal radius (suitably defined; see Section \ref{sec:rt} below) but outside the direct capture radius as TDEs.}, and this effect can be severe for SMBHs with masses such that the tidal radius is comparable to the direct capture radius; as discussed below, this SMBH mass is $\sim 3\times 10^{7}M_{\odot}$ for solar-like stars, and is substantially smaller for less massive stars, for a Schwarzschild black hole (see Equation \ref{Mdccond} and the discussion thereof; see also Section \ref{sec:tderate} below). 

To explore these effects we consider the statistical distribution of pericentre distances of stars encountering SMBHs. In Section \ref{sec:boltzmann} we briefly discuss the relativistic Boltzmann equation, and time-steady solutions to it, and our general assumptions about the nature of stars that are scattered into the tidal disruption (or direct capture) sphere from large distances. We then go on to analyze the distribution of pericenter distances in the Schwarzschild metric (Section \ref{sec:schwarzschild}) and the Kerr metric (Section \ref{sec:kerr}), assuming that the distribution function is isotropic at sufficiently large distances from the SMBH and that it satisfies the Boltzmann equation. One could argue that the analysis of Section \ref{sec:schwarzschild} that focuses on Schwarzschild SMBHs is unnecessary, because we consider the Kerr metric -- of which the Schwarzschild metric is just a special case -- in Section \ref{sec:kerr}. However, solutions with zero spin and that possess complete angular symmetry have a particularly simple, analytic solution for the distribution of pericenter distances and the number of tidally disrupted stars (see Equations \ref{frp} and \ref{Ntde}), both of which provide checks on the more general results in Section \ref{sec:kerr}. In Section \ref{sec:discussion} we provide discussion and implications of our analysis, particularly with respect to the predicted rates of TDEs and their dependence on SMBH spin (Section \ref{sec:tderate}), and the definition of the tidal radius and the generation of \emph{observable} TDEs, i.e., those that likely produce enough electromagnetic emission to be detected to cosmological distances (Section \ref{sec:rt}). We summarize and conclude in Section \ref{sec:summary}. 

\section{Basic assumptions}
\label{sec:boltzmann}
We assume that the distribution of stars in the core of a galaxy is one that satisfies the Boltzmann equation and is therefore approximately collisionless; this is a good approximation for times shorter than the relaxation timescale of the galaxy, assuming that this timescale is long compared to the dynamical time of the stars. Over the relaxation timescale stars will gravitationally interact and induce time dependence owing to the scattering of stars into the region of parameter space that brings them within the tidal disruption (or direct capture) radius of the SMBH, that region of parameter space known as the ``loss cone,'' which has been the focus of many studies (e.g., \citealt{frank76, lightman77, cohn78}). A steady state is reached if stars can repopulate the loss cone fast enough to maintain a constant rate of consumption by the SMBH, and collisions yield no net change to the distribution function, and the Boltzmann equation is again approximately satisfied. 

The relativistic Boltzmann equation for the distribution function $f(x^{\mu},\dot{x}^{\mu})$, where $x^{\mu}$ is the position four-vector, is (e.g., \citealt{mihalas84}) 

\begin{equation}
\frac{d x^{\mu}}{d \tau}\frac{\partial f}{\partial x^{\mu}}+\frac{d \dot{x}^{\mu}}{d \tau}\frac{\partial f}{\partial \dot{x}^{\mu}} = 0. \label{boltzmann}
\end{equation}
Here dots denote differentiation with respect to proper time, Greek indices range from 0 -- 3, and repeated upper and lower indices imply summation. The additional constraint that the four-velocity satisfy $g_{\mu\nu}\dot{x}^{\mu}\dot{x}^{\nu} = -1$ must also be imposed, where $g_{\mu\nu}$ is the metric and is assumed to be dominated by the SMBH at the radii under consideration. Individual particle orbits obey conservation laws of the form 

\begin{equation}
\frac{d}{d\tau}\left[K\left(x^{\mu},\dot{x}^{\mu}\right)\right] = \frac{d x^{\mu}}{d \tau}\frac{\partial K}{\partial x^{\mu}}+\frac{d \dot{x}^{\mu}}{d \tau}\frac{\partial K}{\partial \dot{x}^{\mu}} = 0, \label{Kcon}
\end{equation}
where $K$ is a constant of the motion, and hence a time-steady distribution function can be any function of the set of $K$'s and satisfy the Boltzmann equation (Jeans's theorem; \citealt{jeans1915}).

As pointed out by \citet{merritt13}, the energy relaxation timescales of massive (and the most luminous) galaxies are sufficiently long that the assumption of a steady-state in energy is not actually warranted, and hence the distribution of stars within the sphere of influence of the SMBH may not follow the Bahcall-Wolf scaling $n \propto r^{-7/4}$ \citep{bahcall76}. In other words, stars are not repopulated in energy as fast as they are injected onto orbits that take them within the tidal disruption (or direct capture) radius of the SMBH. On the other hand, the angular momentum relaxation timescale specific to the range of stars that come within the tidal radius is much shorter, and the assumption of satisfying the Boltzmann equation in terms of angular momentum is more justified \citep{merritt13}. 

As such and for the remainder of what follows we will assume that the distribution function is primarily determined by its dependence on angular momentum, and that the majority of stars are scattered from such large distances that the binding energy can be taken to be zero (relativistically, this means that the energy equals the rest-mass energy in the limit that the body is infinitely far from the SMBH) and that the phase space is isotropically populated at large distances. This assumption is tantamount to the statement that stars come from the region of parameter space where they enter into and out of the loss cone on a per-orbit basis, i.e., where the loss cone is ``full,'' and does not account for the stars that slowly (relative to the orbital time) diffuse in energy and angular momentum across the loss cone boundary (the latter being the boundary to the ``empty'' region of the loss cone). A substantial fraction of stars always comes from the full loss cone region, and it dominates for relatively low-mass galaxies \citep{merritt13}. We return to a discussion of the empty loss cone, and whether or not it actually contributes to TDEs that can actually be detected, i.e., produce copious amounts of luminous emission, in Section \ref{sec:rt}.

In the next section we consider solutions to the Boltzmann equation that are spherically symmetric in the Schwarzschild metric, for which only the energy and the magnitude of the angular momentum are relevant as concerns the pericenter distances of tidally disrupted stars, and to the Kerr metric in Section \ref{sec:kerr} where the rotation of the SMBH implies that, even if the distribution function is isotropic at large distances from the SMBH, at small radii the projection of the angular momentum onto the spin axis of the SMBH is also important for stars that are tidally disrupted. We adopt units with $G = M = c = 1$ unless otherwise noted.

\section{The pericenter distribution in the Schwarzschild metric}
\label{sec:schwarzschild}
The Schwarzschild metric is

\begin{equation}
ds^2 = -\left(1-\frac{2}{r}\right)dt^2+\left(1-\frac{2}{r}\right)^{-1}dr^2+r^2 d\Omega^2,
\end{equation}
where $d\Omega^2 = d\theta^2+\sin^2\theta d\phi^2$, and the symmetries of the metric yield three conservation laws for individual-particle orbits:

\begin{equation}
J^2 = r^2\left(r^2\dot{\theta}^2+r^2\sin^2\theta\dot{\phi}^2\right), \label{Js}
\end{equation}
\begin{equation}
\epsilon-1 = \frac{1}{2}\dot{r}^2+\frac{1}{2}\frac{J^2}{r^2}-\frac{1}{r}-\frac{J^2}{r^3},  \label{Es}
\end{equation}
\begin{equation}
\ell = r^2\sin^2\theta\dot{\phi}. \label{ells}
\end{equation}
Here dots denote differentiation with respect to proper time, and these conserved quantities represent the total angular momentum (squared; $J^2$), the component of the angular momentum perpendicular to the $\theta = 0$ axis ($\ell$), and the total relativistic binding energy ($\epsilon$). The pericenter distance $r_{\rm p}$ is obtained by setting $\dot{r} = 0$ in Equation \eqref{Es}; as motivated in the previous subsection, assuming that stars are primarily scattered onto low-angular momentum orbits from large distances and have $\epsilon = 1$, the pericenter distance is related to $J^2$ via

\begin{equation}
J^2 = \frac{2r_{\rm p}^2}{r_{\rm p}-2}. \label{rpJs}
\end{equation} 
Since $\ell$ does not have any bearing on the pericenter distance for this metric, we can marginalize over this variable without loss of generality. 

Equation \eqref{rpJs} has a relative minimum at $J^2 = 16$ and $r_{\rm p} = 4$, which corresponds to the direct capture radius; for $J^2 < 16$ we have $r_{\rm p} = 0$. The solution for the pericenter distance $r_{\rm p}$ as a function of $J^2$, as can be determined by inverting Equation \eqref{rpJs}, is therefore

\begin{equation}
r_{\rm p}(J^2) = 
\begin{cases}
\frac{J^2}{4}\left(1+\sqrt{1-\frac{16}{J^2}}\right) & \textrm{ for } J^2 \ge 16 \\
0 & \textrm{ for } J^2 < 16. \label{rmp}
\end{cases}
\end{equation}
We are interested in stars that reach pericenter distances $r_{\rm p} < r_{\rm t}$, where $r_{\rm t}$ is the tidal radius (suitably defined; see below), and hence have $J^2 < J_{\rm t}^2 = 2r_{\rm t}^2/(r_{\rm t}-2)$. If stars at large distances from the SMBH can be approximated as isotropic in position and velocity space, then it is straightforward to show that the distribution of $J^2$ is uniform in the limit of small-$J^2$ relative to $r^2v^2$ (see Appendix \ref{sec:appendix}). Denoting the distribution function of $J^2$ by $f_{\rm J^2}$, the distribution function of $r_{\rm p}$ is

\begin{equation}
f_{\rm r_{\rm p}}(r_{\rm p}) = \int \delta\left[r_{\rm p}-r_{\rm p}(J^2)\right]f_{\rm J^2}(J^2)dJ^2, \label{rmp2}
\end{equation}
where $r_{\rm p}(J^2)$ is given by Equation \eqref{rmp}. If the total number of stars with $J^2 < J_{\rm t}^2$ is defined as $N$, then straightforward manipulation of Equation \eqref{rmp2} with $f_{\rm J^2}$ equal to a constant shows that the distribution function of $r_{\rm p}$ is

\begin{equation}
f_{\rm r_{\rm p}}(r_{\rm p}) = N \times 
\begin{cases}
\frac{8\left(r_{\rm t}-2\right)}{r_{\rm t}^2}\delta(r_{\rm p}) & \textrm{ for } r_{\rm p} < 4 \\
\frac{r_{\rm t}-2}{r_{\rm t}^2}\frac{r_{\rm p}\left(r_{\rm p}-4\right)}{\left(r_{\rm p}-2\right)^2} & \textrm{ for } 4 < r_{\rm p} < r_{\rm t}
\end{cases} \label{frp}
\end{equation}

\begin{figure*}[htbp] 
   \centering
   \includegraphics[width=0.495\textwidth]{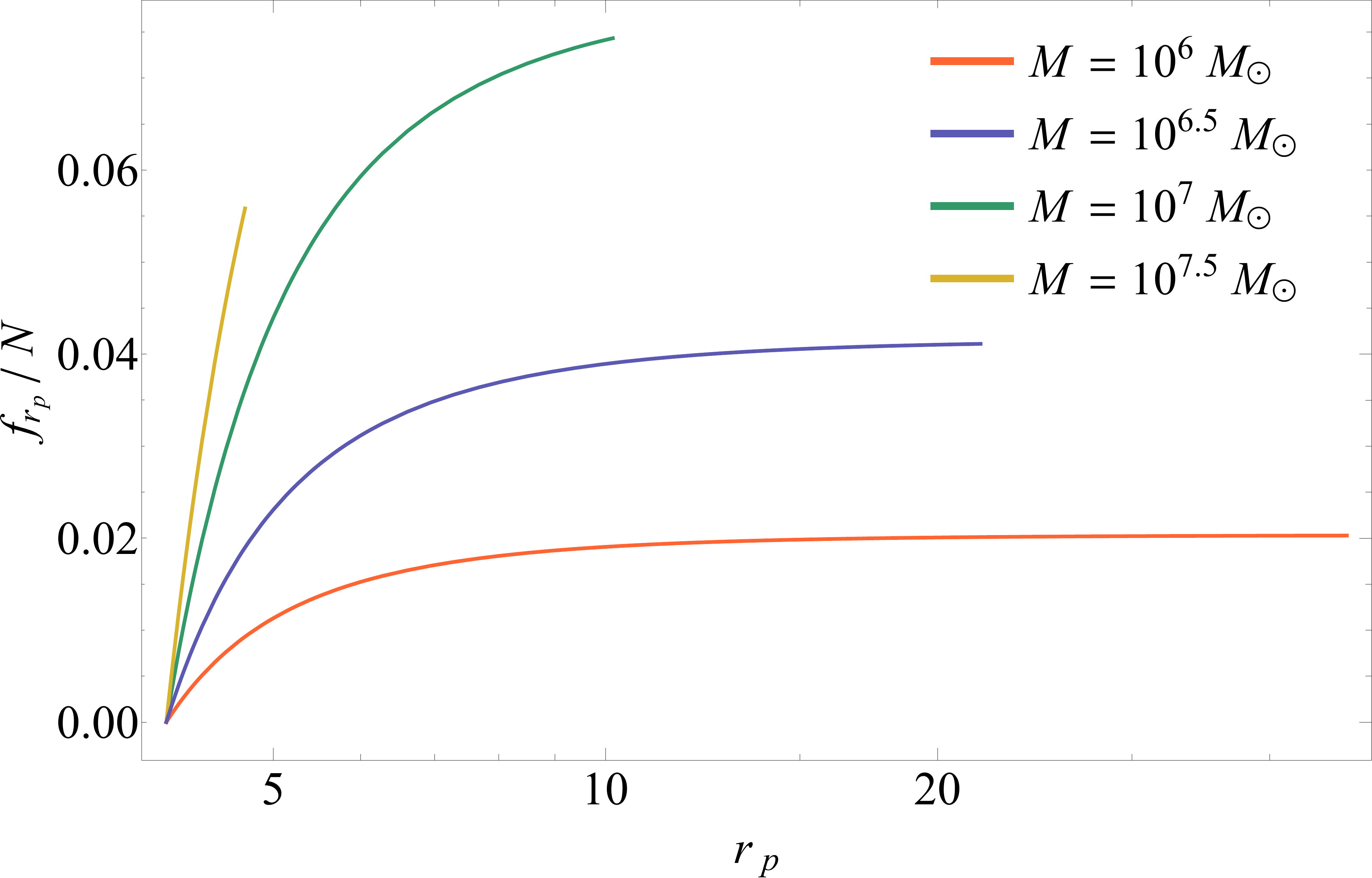} 
  \includegraphics[width=0.495\textwidth]{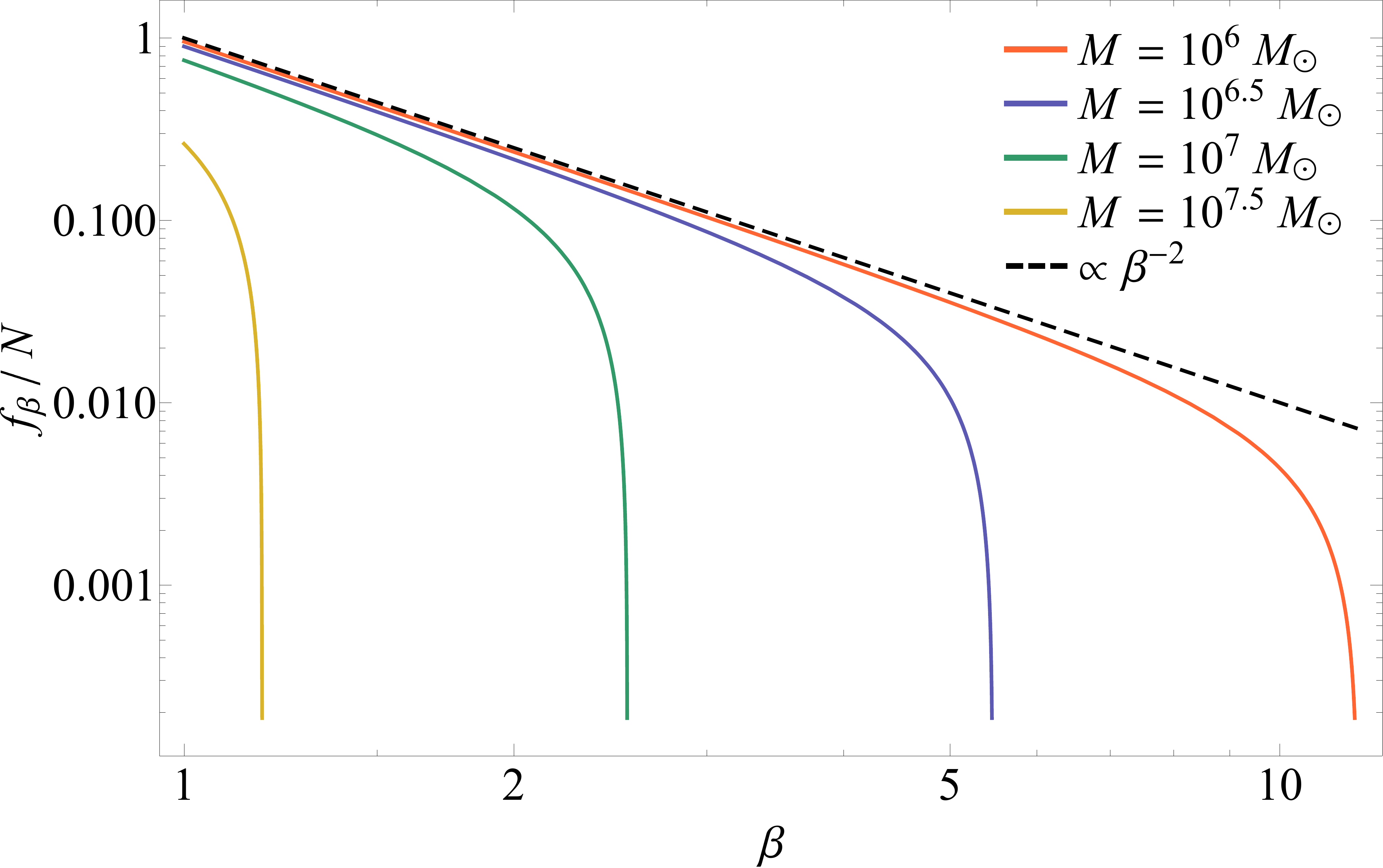} 
   \caption{The distribution function of the pericenter distance (left) and the $\beta$ of the encounter (right), where $\beta = r_{\rm t}/r_{\rm p}$, normalized by $N$, which is the total number of stars scattered within distances $\le r_{\rm t}$. In the left panel the value of $r_{\rm p}$ ranges from the direct capture radius $r_{\rm p}=4$ to the tidal radius $r_{\rm p}=r_{\rm t}$. Note that the integral under each of these curves is less than $N$, the number of stars that enter the tidal radius, because the fraction of stars with $r_{\rm p} < 4$ is directly captured. The Newtonian result is plotted in the right panel as a dashed line with $f_{\beta} \propto \beta^{-2}$; on the left panel the Newtonian result would be a horizontal line with $f_{r_{\rm p}} = $ constant, to which the solutions clearly tend in the large-$r_{\rm p}$ limit.}
   \label{fig:frp}
\end{figure*}

Figure \ref{fig:frp} shows the distribution function for $r_{\rm p} > 4$, given by Equation \eqref{frp}. The SMBH mass for each curve, shown in the legend, is fixed by setting the tidal radius in physical units to 

\begin{equation}
r_{\rm t} = R_{\star}\left(M/M_{\star}\right)^{1/3}, \label{rtdef}
\end{equation}
which is the canonical definition and that we motivate in more detail in Section \ref{sec:rt} below, and using solar values for the star. For a $10^6M_{\odot}$ SMBH, we have $r_{\rm t} \simeq 47 \equiv r_{\rm t, \odot}$ (in gravitational units), which is where the orange curve in Figure \ref{fig:frp} ends. Further note that the dependence on $r_{\rm t}$ enters only through the normalization, which is apparent from Equation \eqref{frp}. Each curve is normalized by the total number of stars that enter the tidal radius, $N$, and hence the fraction of TDEs (i.e., the area under each curve) decreases as the SMBH mass increases. Because it is customarily referred to in the TDE literature, we can also calculate the distribution function of $\beta \equiv r_{\rm t}/r_{\rm p}$, which from Equation \eqref{frp} is

\begin{equation}
f_{\beta}(\beta) = N \times
\begin{cases}
\frac{8\left(r_{\rm t}-2\right)}{r_{\rm t}^2}\delta\left[\beta-\infty\right] & \textrm{ for } \beta > \frac{r_{\rm t}}{4} \\ 
\frac{r_{\rm t}-2}{\beta^2}\frac{r_{\rm t}-4\beta}{\left(r_{\rm t}-2\beta\right)^2} & \textrm{ for } 1 < \beta < \frac{r_{\rm t}}{4}.
\end{cases} \label{fbeta}
\end{equation}
This distribution function is shown in the right-hand panel of Figure \ref{fig:frp} for the same SMBH masses. The Newtonian limit of $f_{\beta} \propto \beta^{-2}$ is shown by the black, dashed line for reference. The presence of the direct capture radius of the SMBH induces a steep falloff in the distribution function near $\beta = r_{\rm t}/4$.

\begin{figure}[htbp] 
   \centering
   \includegraphics[width=0.47\textwidth]{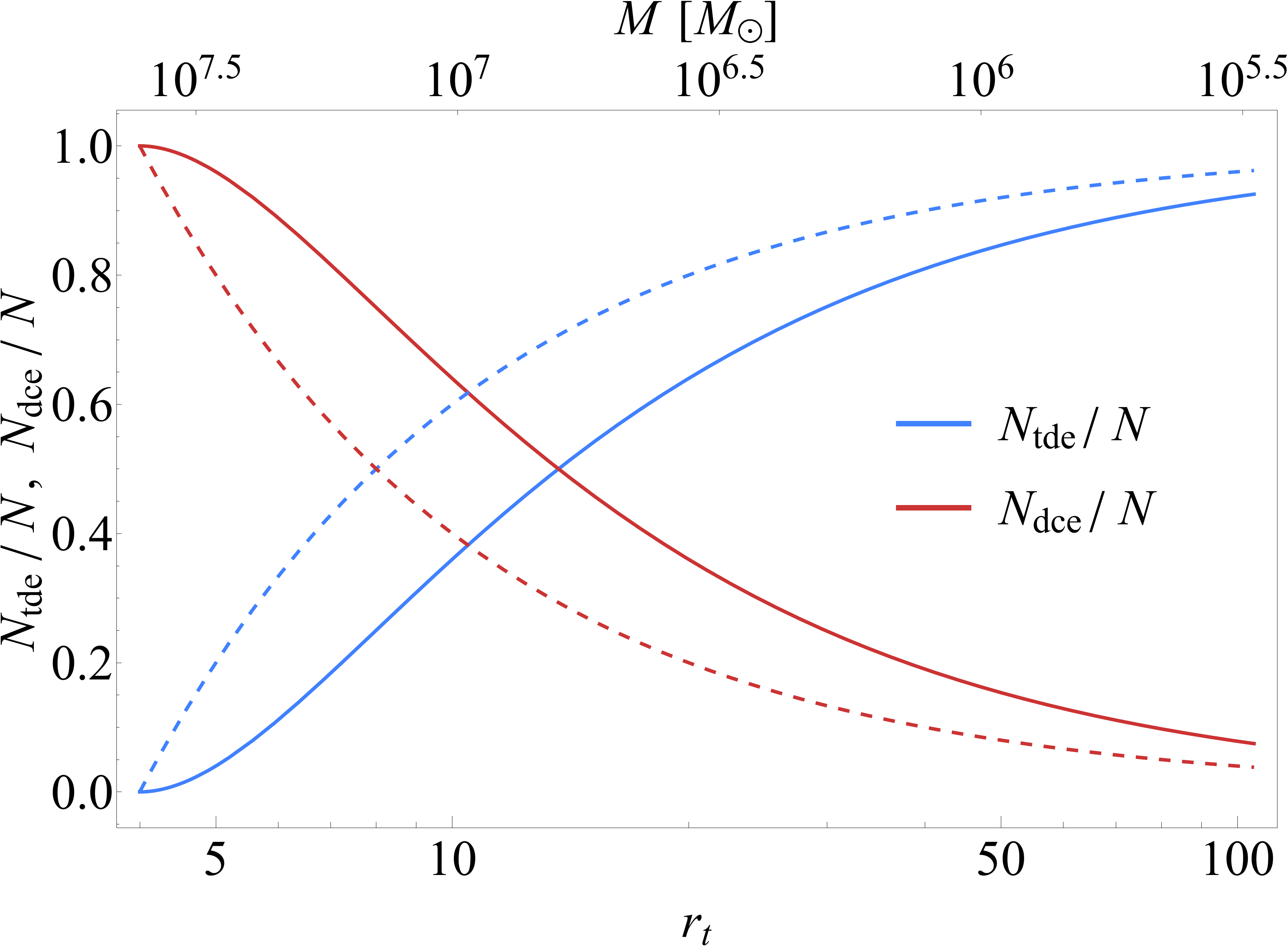} 
   \caption{The fraction of TDEs (blue) and DCEs (red) as a function of the tidal radius {\bf (in gravitational radii; $r_{\rm g} = GM/c^2$)} of the SMBH. The solid lines account for relativistic effects (Equations \ref{Ntde} and \ref{Ndc}), while the dashed lines are in the Newtonian limit (Equation \ref{NtdeNewt}). The relation between the SMBH mass (top axis) and the tidal radius is given in Equation \eqref{rtdef} and is for a solar-like star.}
   \label{fig:Ntde_Ndc}
\end{figure}

Integrating Equation \eqref{frp} shows that the number of TDEs, equal to the integral of $f_{\rm r_{\rm p}}$ from $r_{\rm p} = 4$ to $r_{\rm t}$, is

\begin{equation}
N_{\rm tde} = \int_4^{r_{\rm t}} f_{\rm r_{\rm p}}(r_{\rm p})\,d r_{\rm p} = \left(1-\frac{4}{r_{\rm t}}\right)^2N, \label{Ntde}
\end{equation}
and the number of directly captured stars, or ``direct capture events (DCEs),'' is ($N-N_{\rm tde}$; as can be verified, this is also the integral of $f_{\rm r_{\rm p}}$ from 0 to $4$ with $f_{\rm r_{\rm p}}$ given by Equation \ref{frp})

\begin{equation}
N_{\rm dce} = N\left\{1-\left(1-\frac{4}{r_{\rm t}}\right)^2\right\}. \label{Ndc}
\end{equation}
Thus, for a SMBH mass of $10^6M_{\odot}$, $\sim 83.7\%$ of encounters yield TDEs (and $\sim 16.3\%$ DCEs), while a SMBH mass of $10^{7.5}M_{\odot} \simeq 3.2\times 10^{7}M_{\odot}$ has a tidal disruption fraction of $2.3\%$ (and a direct capture fraction of $\sim 97.7\%$). By comparison, the Newtonian distribution of pericenters is uniform in $r_{\rm p}$, which gives

\begin{equation}
N_{\rm tde,\, Newt} = \frac{N}{r_{\rm t}}\left(r_{\rm t}-4\right), \quad N_{\rm dce, \, Newt} = \frac{4N}{r_{\rm t}}. \label{NtdeNewt}
\end{equation}
Figure \ref{fig:Ntde_Ndc} shows the relative fraction of tidally disrupted (blue) and directly captured (red) stars with relativistic effects included (solid) and in the Newtonian regime (dashed), and we used Equation \eqref{rtdef} to plot these ratios as a function of SMBH mass on the top axis. Even though the curves equal one another at $r_{\rm t} = 4$ and in the limit that $r_{\rm t}\rightarrow \infty$, there are substantial differences between the two; the difference is maximized at $0.25$ when $r_{\rm t} = 8$ ($M \simeq 1.4\times10^{7}M_{\odot}$) where the Newtonian approximation predicts that 50\% of stars are tidally disrupted, whereas the relativistic value is $25\%$. Thus, the Newtonian approximation can significantly over-predict the number of TDEs.

The total number of TDEs with $\beta > \beta_{\rm min}$ is

\begin{equation}
\frac{1}{N_{\rm tde}} \int_{r_{\rm t}/4}^{\beta_{\rm min}}f_{\beta}(\tilde{\beta})d\tilde{\beta} = \frac{\left(1-\frac{2}{r_{\rm t}}\right)\left(1-\frac{4\beta_{\rm min}}{r_{\rm t}}\right)^2}{\beta_{\rm min}\left(1-\frac{2\beta_{\rm min}}{r_{\rm t}}\right)\left(1-\frac{4}{r_{\rm t}}\right)^2}. \label{betacdf}
\end{equation}
For example, the canonical tidal radius for a $10^{6}M_{\odot}$ SMBH is (see Equation \ref{rtdef}) $r_{\rm t} \simeq 47$ and the total fraction of encounters that have $\beta > 6$ (i.e., $\beta_{\rm min} = 6$) is $\sim 6.1\%$. On the other hand, the Newtonian limit is 

\begin{equation}
F_{\beta,\rm Newtonian} = \frac{1}{\beta}\frac{1-\frac{4\beta}{r_{\rm t}}}{1-\frac{4}{r_{\rm t}}}.
\end{equation}
With $r_{\rm t} =r_{\rm t, \odot} \simeq 47$, this predicts that $\simeq 8.9\%$ of encounters have $\beta > 6$ and produce TDEs, which is a factor of $\sim 1.5$ larger than the true (relativistic) value given above. The disagreement between the Newtonian and relativistic values is because relativistic gravity draws stars to smaller radii for a given $J^2$ compared to the Newtonian pericenter. This is most apparent by comparing Equations \eqref{Ntde}, \eqref{Ndc}, and \eqref{NtdeNewt}.

Equation \eqref{Ntde} shows that when $r_{\rm t} = 4$, the number of TDEs equals zero and all stars are directly captured (obviously this is true for $r_{\rm t} \le 4$). We must therefore have, in physical units, 

\begin{equation}
r_{\rm t} > \frac{4GM}{c^2} \label{rtlim}
\end{equation}
for a spinless SMBH to produce \emph{any} observable disruptions. This expression for the direct capture radius of a Schwarzschild SMBH is consistent with, e.g., \citet{zeldovich71, bardeen72, beloborodov92, will12} (note that \citealt{zeldovich71, beloborodov92} equate the gravitational radius $r_{\rm g}$ to what is more commonly referred to as the Schwarzschild radius, i.e., they let $r_{\rm g} = 2GM/c^2$). This inequality disagrees with the more common claim of $r_{\rm t}>2GM/c^2$ (see Section \ref{sec:tderate} below for additional discussion). 

With the standard definition of the tidal radius given in Equation \eqref{rtdef}, Equation \eqref{rtlim} implies that
\begin{equation}
\frac{M}{M_{\star}} \le \frac{1}{8}\left(\frac{R_{\star}}{GM_{\star}/c^2}\right)^{3/2}  \simeq 4.0\times 10^{7}\left(\frac{R_{\star}}{R_{\odot}}\right)^{3/2}\left(\frac{M_{\star}}{M_{\odot}}\right)^{-3/2} \label{Mdccond}
\end{equation}
if the (Schwarzschild) SMBH is to produce observable TDEs\footnote{Note that Equation \eqref{Mdccond} can also be written as ${\bar \rho}_\star < {\bar \rho}_\bullet$, where ${\bar \rho}_\star = M_\star/R_\star^3$ and ${\bar \rho}_\bullet = M_\bullet/r_{\rm dc}^3$ with $r_{\rm dc} = 4GM/c^2$.}. Here we have reintroduced factors of $G$, $c$, and $M$ for clarity. We discuss the implications of this result in the context of the rates of TDEs in Section \ref{sec:tderate} below. 

\section{The pericenter Distribution in the Kerr metric}
\label{sec:kerr}
In the Kerr metric written in Boyer-Lindquist coordinates, the three conserved quantities are

\begin{equation}
\left(\left(r^2+a^2\cos^2\theta\right)\dot{\theta}\right)^2+\frac{\ell^2}{\sin^2\theta}-a^2\left(1-\epsilon^2\right)\sin^2\theta = J^2, \label{kcon}
\end{equation}
\begin{equation}
\left(1-\frac{2r}{r^2+a^2\cos^2\theta}\right)\dot{t}+\frac{2 a r \sin^2\theta}{r^2+a^2\cos^2\theta}\dot{\phi} = \epsilon,
\end{equation}
\begin{multline}
\frac{\left(r^2+a^2\right)^2\sin^2\theta-a^2\left(r^2-2 r+a^2\right)\sin^4\theta}{r^2+a^2\cos^2\theta}\dot{\phi} \\
-\frac{2a r\sin^2\theta}{r^2+a^2\cos^2\theta}\dot{t} = \ell
\end{multline}
These are the relativistic generalizations of the square of the Newtonian total angular momentum, the specific energy, and the projection of the angular momentum onto the spin axis of the SMBH, to which they manifestly reduce in the limit of large $r$. The conservation of the norm of the four-velocity also gives $g_{\mu\nu}\dot{x}^{\mu}\dot{x}^{\nu} = -1$, which yields a fourth, nonlinear equation relating the first temporal derivatives of the coordinates. In the limit that the particles are on zero-energy orbits, so $\epsilon \equiv 1$ in the previous equations, we can show that these four equations can be combined to give the following relationship among $J^2$, $\ell$, and the pericenter distance $r_{\rm p}$ (which, as for the Schwarzschild case, is obtained by setting $\dot{r}=0$; note that $\dot{r}$ explicitly appears in the condition $g_{\mu\nu}\dot{x}^{\mu}\dot{x}^{\nu} = -1$):

\begin{equation}
J^2 = \frac{2r_{\rm p}^3+2a^2r_{\rm p}-4ar_{\rm p}\ell+a^2\ell^2}{r_{\rm p}^2-2r_{\rm p}+a^2}. \label{kofrp}
\end{equation}
If $a \neq 0$, $r_{\rm p}$ depends both on the total angular momentum and the projection of the angular momentum onto the spin axis of the SMBH, $\ell$, because the spin of the SMBH breaks the isotropy of the spacetime. If $a = 0$, Equation \eqref{kofrp} is clearly identical to Equation \eqref{rpJs}.

Stars that produce TDEs must have a pericenter distance $r_{\rm p}$ outside of the direct capture radius\footnote{In the limit that the star is extremely highly elongated at the time it reaches $\sim$ the direct capture radius, which occurs for very high-$\beta$ and low-mass SMBHs, it is possible for the material to promptly self-intersect owing to the extreme advance of periapsis ($\gtrsim 2\pi$), which would likely produce a very short-lived, electromagnetic outburst \citep{darbha19}. However, it is likely that such an outburst would not resemble a traditional TDE.}. The direct capture condition that relates $J^2$ and $\ell$ can be determined by solving Equation \eqref{kofrp} for $r_{\rm p}$ and setting the radical in the solution of the cubic to zero; for completeness, this condition is

\begin{multline}
\left(18 a^2 \left(2 J^2-3 \ell ^2\right)+36 a J^2 \ell +\left(J^2-18\right) J^4\right)^2 \\ 
+\left(12
   \left(a^2-2 a \ell +J^2\right)-J^4\right)^3 = 0. \label{dccond}
\end{multline}
This can be written as a quartic in $\ell$ that can be solved analytically for $\ell(J^2)$, but the solution is not enlightening and, in practice, is more easily solved numerically (the left panel of Figure \ref{fig:dccurves} in Appendix \ref{sec:appendix} shows the numerical solution for the direct capture condition, i.e., the direct capture curves, for a range of $a$). See \citet{will17} for an approximate, analytic expression (their Section A). 

In contrast to the Schwarzschild case, the direct capture condition in the Kerr metric is algebraically more complex and dependent on two variables ($\ell$ and $J^2$), which makes the analysis of (e.g.) the distribution of pericenter distances of tidally disrupted stars much more involved. The joint probability distribution function $f(\ell, J^2)$ itself is also not trivial, even for the case of a spherically symmetric distribution of stars at large distances from the SMBH, because of the fact that $\ell$ and $J^2$ are not independent. To maintain the readability of the paper, here we focus only on the results and we defer the calculation of the joint probability distribution function $f(\ell,J^2)$ to Appendix \ref{sec:appendix}, and the formalism and analysis that exploits this distribution function to infer the properties of disrupted stars -- such as the distribution of pericenter distances and the fraction of TDEs -- to Appendix \ref{sec:appB}. 

\begin{figure*}[htbp] 
   \centering
   \includegraphics[width=0.495\textwidth]{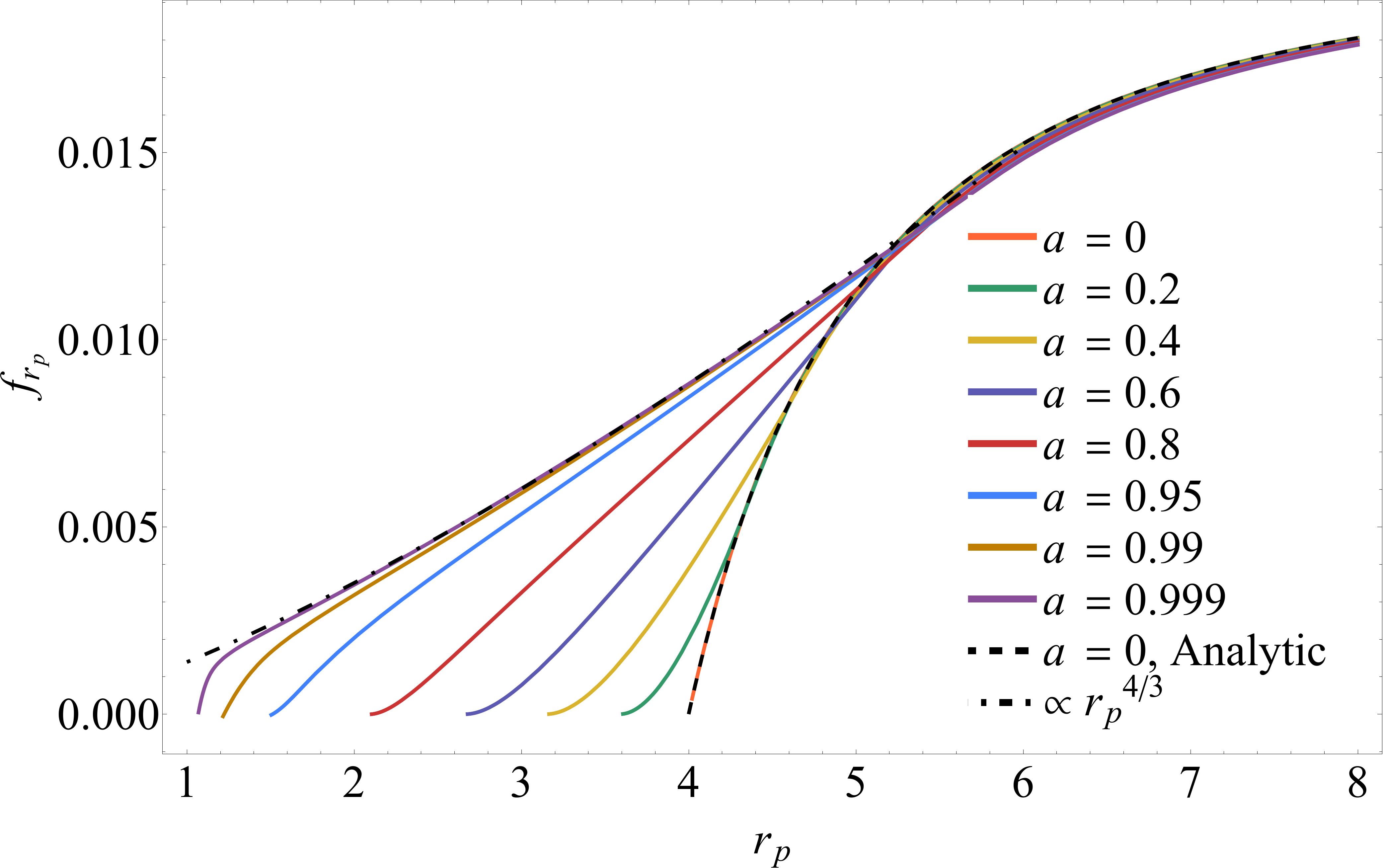} 
   \includegraphics[width=0.495\textwidth]{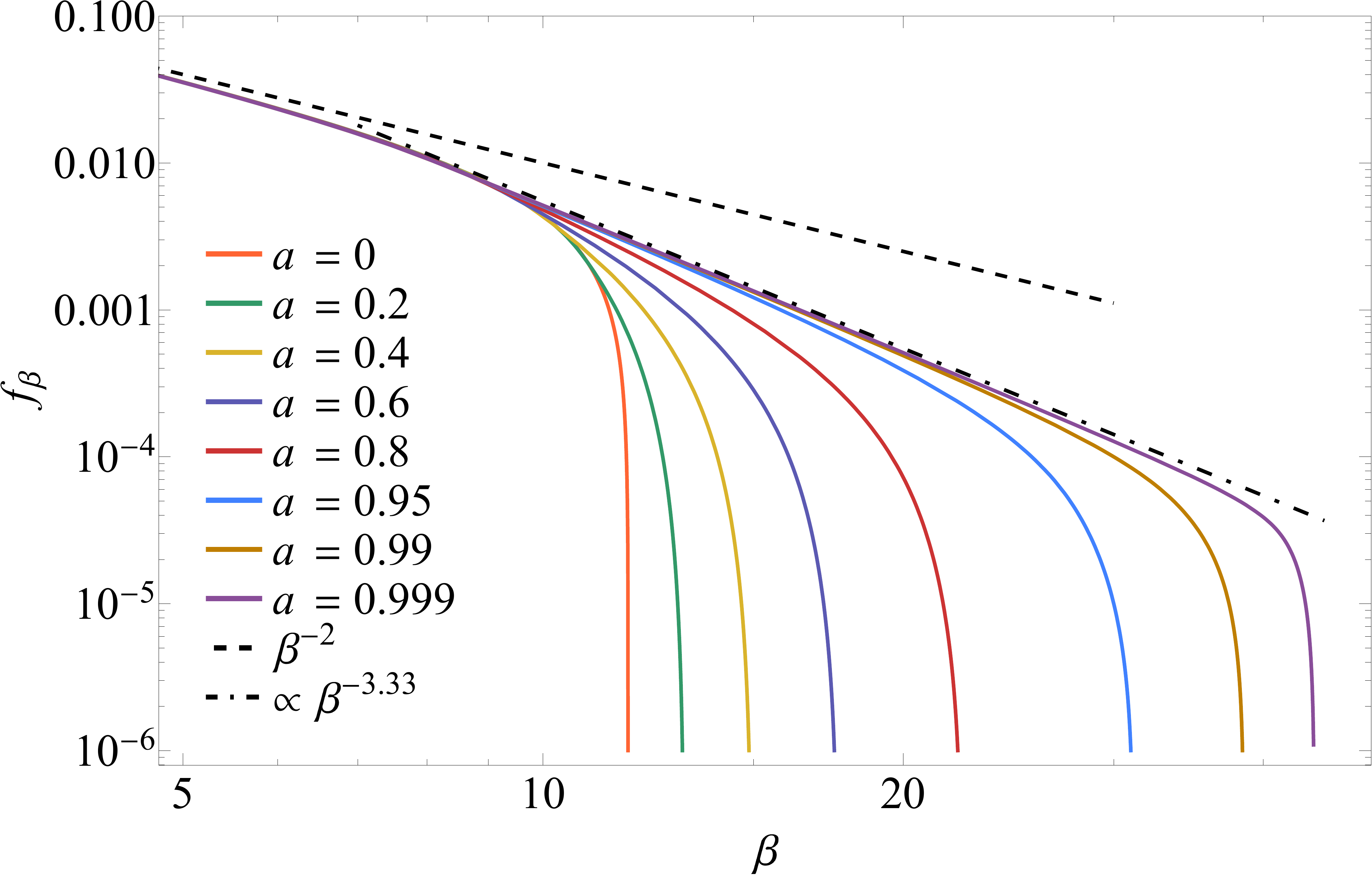} 
   \caption{The distribution function of $r_{\rm p}$ (left) and $\beta$ (right) for a tidal radius of $r_{\rm t} = r_{\rm t, \odot} \simeq 47$ for the SMBH spins in the legend. The black, dashed curve in the left-hand panel is the analytic solution for a Schwarzschild SMBH calculated in Section \ref{sec:schwarzschild}, and is effectively a check on the analysis. In the right panel, the dashed curve gives $f_{\beta} = \beta^{-2}$, which is the result in the Newtonian limit, while the dot-dashed curve shows $f_{\beta} \propto \beta^{-10/3}$. The locations at which the distribution function goes to zero demarcate the smallest possible value of $r_{\rm p}$ on the direct capture curve, which approaches 1 in the limit that $a \rightarrow 1$. }
   \label{fig:frp_kerr}
\end{figure*}

Figure \ref{fig:frp_kerr} gives the probability distribution function of the pericenter distance $r_{\rm p}$ (left) and $\beta = r_{\rm t}/r_{\rm p}$ (right) with $r_{\rm t, \odot} \simeq 47$ and the SMBH spins in the legend. To calculate these solutions we numerically integrated the distribution function $f(\ell,J^2)$ over the region of parameter space that produces TDEs for a finely sampled set of $r_{\rm p}$, interpolated the solution, and differentiated with respect to $r_{\rm p}$ (see Appendices \ref{sec:appendix} and \ref{sec:appB}). These solutions are normalized by the total number of stars that have $r_{\rm p} < r_{\rm t}$, i.e., the integral over of $f_{\rm r_{\rm p}}$ over all $r_{\rm p}$ equals the fraction of TDEs relative to the total number that have $r_{\rm p} < r_{\rm t}$ and hence is always less than one. Independent of spin, the curves approximately equal one another at large $r_{\rm p}$ or small $\beta$ and approximate the analytic solution in the Schwarzschild (spin zero) limit, which is shown by the black, dashed curve in the left panel (and effectively serves as a check on this method). For small $r_{\rm p}$ (large $\beta)$, the distribution function shows a marked difference, and extends to the minimum possible pericenter distance along the direct capture curve that is reachable by the star; from Equation \eqref{rmin}, this minimum possible distance extends to 1 in the limit of maximal spin, which is reflected in this plot. As for the Schwarzschild case, the distribution function in terms of $r_{\rm p}$ depends only on $r_{\rm t}$ through its normalization, and otherwise the only dependence is on $r_{\rm p}$ for a given $a$. In the limit of maximal spin, the right plot of Figure \ref{fig:frp_kerr} shows that the distribution function of $\beta$ at large $\beta$ is $f_{\beta} \propto \beta^{-10/3}$, which implies that the distribution function of $f_{\rm r_{\rm p}} \propto r_{\rm p}^{4/3}$ for small $r_{\rm p}$ and rapidly rotating holes; this behavior is shown by the black, dot-dashed curve in the left-hand panel of Figure \ref{fig:frp_kerr}. 

\begin{figure}[htbp] 
   \centering
  \includegraphics[width=0.475\textwidth]{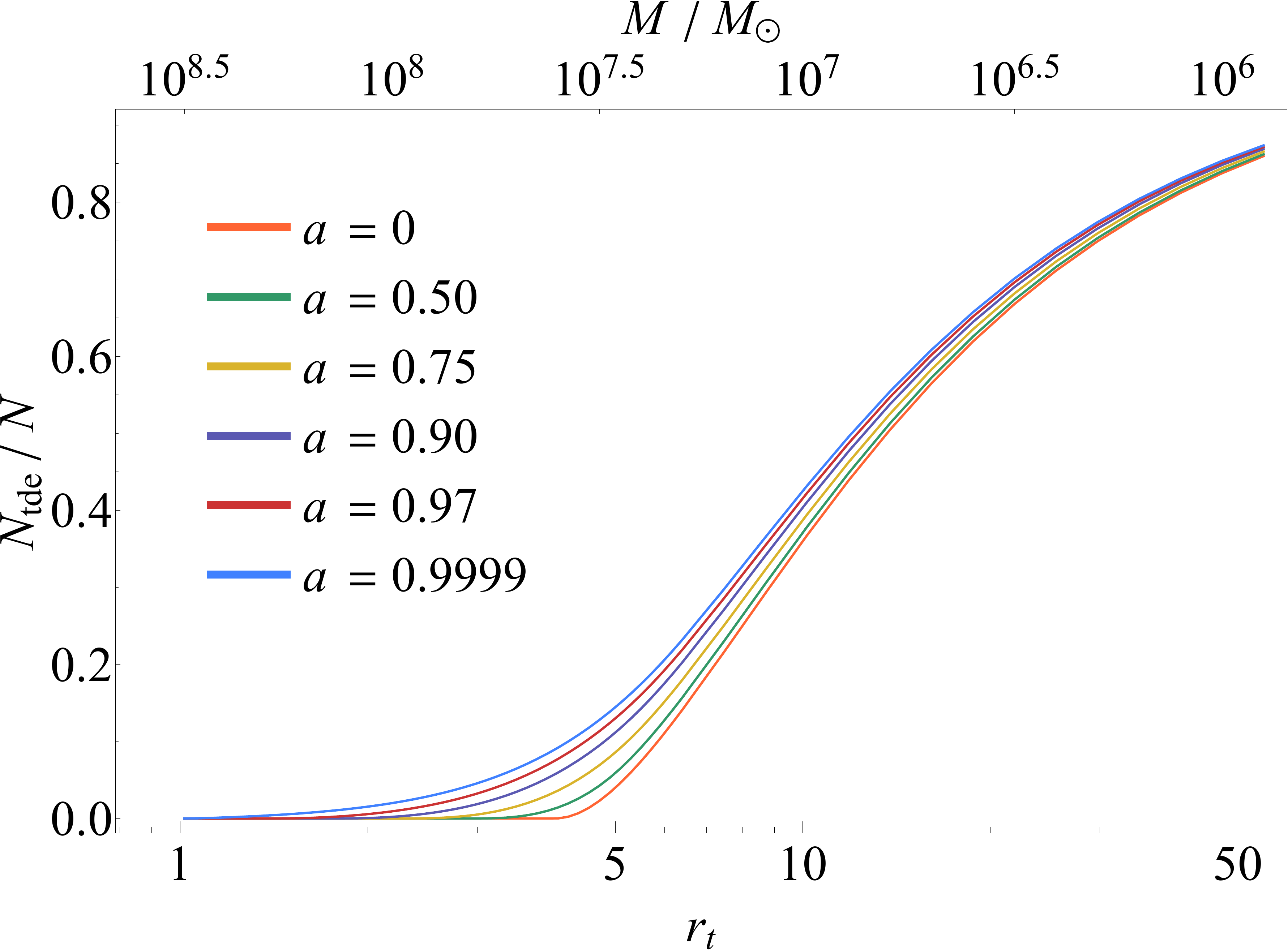} 
   \caption{The number of TDEs relative to the total number of stars scattered into the loss cone as a function of the tidal radius {\bf (in units of gravitational radii)} for the SMBH spins shown in the legend. It is clear that SMBH spin only matters as concerns the rate of TDEs if the SMBH is rapidly rotating, with $a \gtrsim 0.75$. The black hole mass (top axis) is derived from the standard definition for the tidal radius given a solar-like star.}
   \label{fig:Ntde_spin}
\end{figure}

\begin{figure}[htbp] 
   \centering
  \includegraphics[width=0.475\textwidth]{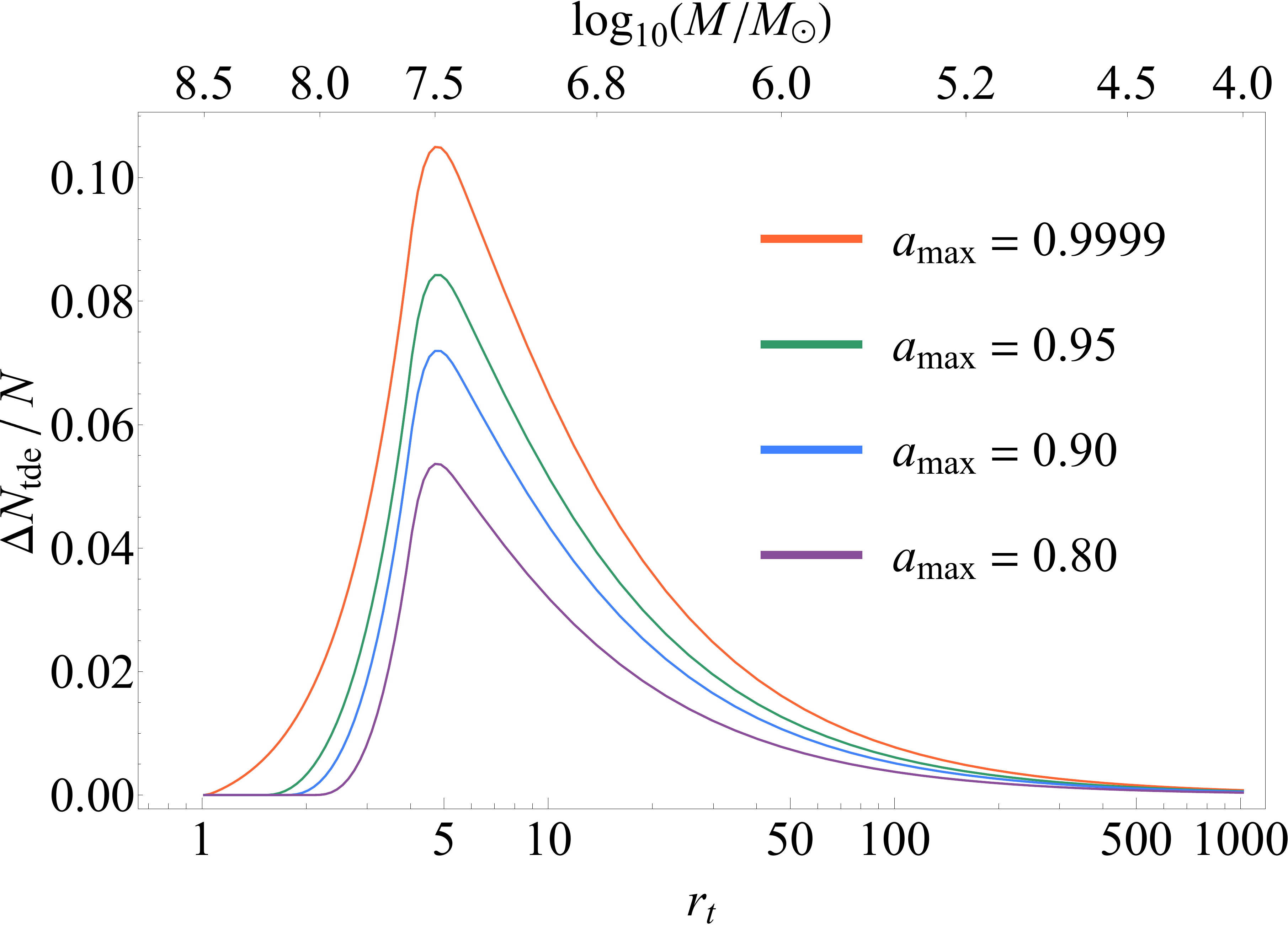} 
   \caption{The difference in the number of TDEs, relative to the total number of stars scattered into the loss cone, between a SMBH with spin given in the legend and a Schwarzschild SMBH. This shows that rapid SMBH spin can generate at most a $\sim 10\%$ difference in the fraction of TDEs, with the maximum difference occurring at a SMBH mass of $\sim 3.2\times 10^{7}M_{\odot}$. }
   \label{fig:DeltaN}
\end{figure}

Figure \ref{fig:Ntde_spin} shows the number of tidally disrupted stars relative to the total number scattered into the loss cone, $N_{\rm tde}/N$, as a function of the tidal radius for the SMBH spins in the legend. The top axis gives the SMBH mass, which is calculated from Equation \eqref{rtdef}. It is apparent from this figure that the spin of the SMBH plays little role in modifying the number of TDEs until the tidal radius is only marginally greater than the gravitational radius, and even then only for rapidly rotating SMBHs with $a \gtrsim 0.75$. To quantify these statements, Figure \ref{fig:DeltaN} shows the difference in the fraction of stars that yield TDEs for a spinning SMBH, with the spin given in the legend, and a Schwarzschild SMBH, as a function of the tidal radius $r_{\rm t}$ (bottom axis) and the SMBH mass (upper axis) assuming solar-like stars. These curves show a peak at a tidal radius of 

\begin{equation}
r_{\rm t, peak} \simeq 4.7,
\end{equation}
and if we adopt our definition of the tidal radius, at a SMBH mass of

\begin{equation}
\frac{M}{M_{\star}} \simeq 3.2\times 10^{7}\left(\frac{R_{\star}}{R_{\odot}}\right)^{3/2}\left(\frac{M_{\star}}{M_{\odot}}\right)^{-3/2}.
\end{equation}
This SMBH mass that would maximize the difference in the number of TDEs is slightly smaller than the one at which the direct capture radius coincides with the tidal radius, being $\sim 4\times 10^{7}M_{\odot}$ for solar-like stars. 

In the next section we discuss the implications of our results.

\section{Discussion and Implications}
\label{sec:discussion}
In the previous two sections we derived the pericenter distribution function of stars scattered into the loss cone of a SMBH for the Schwarzschild (Section \ref{sec:schwarzschild}) and Kerr (Section \ref{sec:kerr}) metrics, and from these the relative number of TDEs to the total number of stars that have small enough angular momentum to come within $r_{\rm t}$ (Figure \ref{fig:Ntde_spin}). Here we discuss the implications of these results in the context of the rates of TDEs.

\subsection{TDE rate suppression and the unimportance of SMBH spin}
\label{sec:tderate}
The rate of TDEs per galaxy is

\begin{multline}
\dot{N}_{\rm tde}(M)\left[\textrm{gal}^{-1}\textrm{ yr}^{-1}\right] \\ 
= \int\dot{N}(M)\frac{N_{\rm tde}(r_{\rm t},a)}{N}f_{\star}(M_{\star}, R_{\star})dM_{\star}dR_{\star}. \label{ndotsch} 
\end{multline}
Here $f_{\star}$ is the distribution function of stellar masses and radii within the galaxy. $\dot{N}$ is the rate at which stars are supplied to the loss cone and is not, in general, just a function of the SMBH mass; however, a number of investigations have found that it grows weakly with decreasing SMBH mass when the $M$-$\sigma$ relation is incorporated (e.g., \citealt{stone16} find $\dot{N} \propto M^{-0.404}$ over the entire galaxy sample they investigated, though they also note that the rate they derive is effectively independent of SMBH mass for $M < 10^{8}M_{\odot}$). If we take Equation (33) from \citet{merritt13} and an $M$-$\sigma$ relation $M\propto \sigma^5$, normalized such that $\sigma = 200$ km s$^{-1}$ corresponds to a SMBH mass of $2\times 10^{8} M_{\odot}$ \citep[e.g.,][]{marsden20}, we find

\begin{equation}
\dot{N} \simeq 4.5\times 10^{-4}\left(\frac{M}{4\times 10^{6}M_{\odot}}\right)^{-0.3}\textrm{ yr}^{-1}. \label{Ndot}
\end{equation}
As highlighted by \citet{merritt13}, the fact that the rate increases with decreasing SMBH mass implies that low-mass SMBHs contribute predominantly to the TDE rate. 

The dominance of the low-mass end becomes even more pronounced when we incorporate the dependence of $N_{\rm tde}/N$ on $r_{\rm t}$. From Figure \ref{fig:Ntde_spin}, the rate of TDEs equals zero for all $r_{\rm t} \le 4$ for a Schwarzschild ($a = 0$) SMBH, but it is strongly suppressed for all $r_{\rm t} \lesssim 10$ effectively independently of the spin of the SMBH. We would expect a substantial reduction in the rate of TDEs when $N_{\rm tde}/N$ falls below $0.1$, or when $r_{\rm t}$ satisfies 

\begin{equation}
r_{\rm t} \lesssim 5\frac{GM}{c^2}, \label{rtcrit}
\end{equation}
independent of the spin parameter; we have included factors of $G$, $M$ and $c$ in this equation for clarity. 

All of this analysis is independent of our definition of the tidal radius\footnote{\label{footnote:1}The exception is if one chooses to invoke a particularly strong dependence of the tidal radius on the spin of the SMBH that is also not proportional to the angular momentum of the star, i.e., one that is not just of the form $\propto a\ell$ such that the increased (or reduced) rate of disruption of stars with prograde angular momenta cancels that of those with retrograde angular momenta. In addition to being unlikely from a physical standpoint, \citet{gafton19} specifically state that spin introduces at most a 1\% effect as concerns the mass stripped from the star in partial disruption.}. If we use the standard definition given in Equation \eqref{rtdef} and let $r_{\rm t} = R_{\star}\left(M/M_{\star}\right)^{1/3}$, then Equation \eqref{rtcrit} implies that the TDE rate should decline substantially once the SMBH mass satisfies, with factors of $G$, $M$, and $c$ included for clarity,

\begin{multline}
\frac{M}{M_{\star}} \le \left(\frac{R_{\star}}{5GM_{\star}/c^2}\right)^{3/2} \\
 \simeq 2.9\times 10^{7}\left(\frac{R_{\star}}{R_{\odot}}\right)^{3/2}\left(\frac{M_{\star}}{M_{\odot}}\right)^{-3/2}. \label{masslimit}
\end{multline}
This limit on the mass of the SMBH is significantly smaller than the value that is often quoted in the literature\footnote{A notable exception is \citet{merritt13}, who stated that tidally disrupted stars must have a \emph{Newtonian} $r_{\rm t} > 8$ (gravitational radii), and hence concluded that SMBHs capable of producing observable TDEs must have a mass that satisfies $M/M_{\star} \lesssim 1.2\times 10^{7}$ -- a factor of $\sim$ 2.5 smaller than the one in Equation \eqref{masslimit}. \citet{merritt13} let $r_{\rm t} = 8$ based on the work of \citet{will12}, who \emph{defined} the Newtonian pericenter by the expression (for a parabolic orbit) $J^2 = 2r_{\rm p}$ with $J^2 = 16$ (see their Equation 17; this same condition has also been employed in, e.g., \citealt{broggi22}). However, the true pericenter distance reached by the star is $r_{\rm p} = 4$ when $J^2 = 16$, and hence the correct limit is given by Equation \eqref{masslimit} (or Equation \ref{Mdccond} when $r_{\rm p} =4$).} and obtained by setting $r_{\rm t} > 2GM/c^2$, which gives $M/M_{\star} \lesssim 1.1\times 10^{8}$ (for a solar-like star; e.g., \citealt{hills75, magorrian99, syer99, wang04, kesden12, stone16, stone19, stone20}; but see \citealt{servin17}).

\begin{figure}[htbp] 
   \centering
    \includegraphics[width=.47\textwidth]{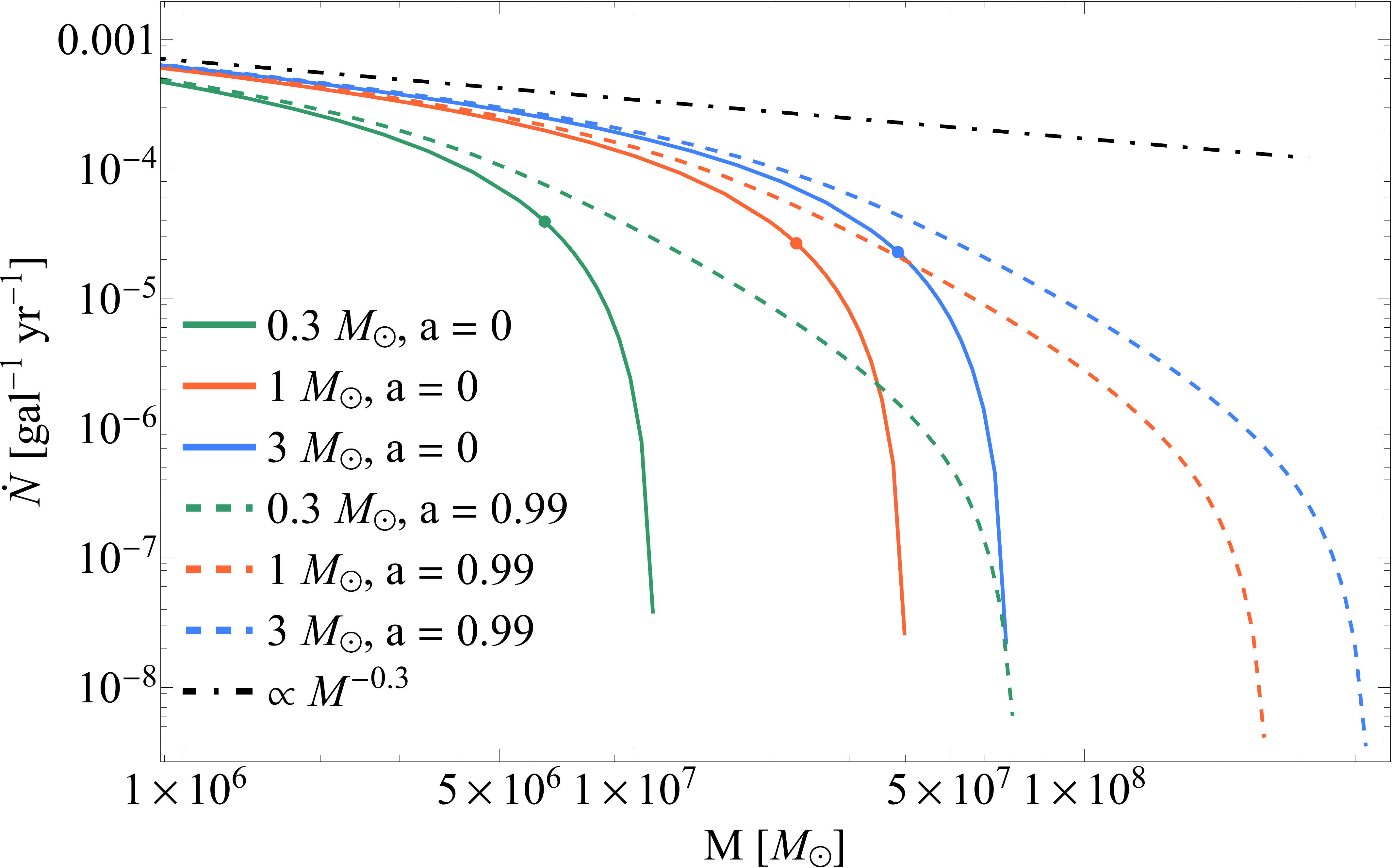} 
     \caption{The TDE rate as a function of SMBH mass for the combinations of SMBH spin and stellar type shown in the legend. The points on each curve show where the rate falls by an order of magnitude relative to the rate at which stars are scattered into the loss cone, where the latter is shown by the black, dot-dashed curve. }
   \label{fig:rates}
\end{figure}

The rate of TDEs for a given galaxy (Equation \ref{ndotsch}) depends on the details of the stellar distribution function, which may display significant variation from galaxy to galaxy, but we can gain insight by considering stellar populations that consist of only one type of star. Figure \ref{fig:rates} shows $\dot{N}_{\rm tde}$ with Equation \eqref{Ndot} for $\dot{N}$ and for $M_{\star} = 0.3$, 1, and 3 $M_{\odot}$. The radius of the $1M_{\odot}$ star is equal to $1R_{\odot}$ by construction, whereas the $0.3M_{\odot}$ and $3M_{\odot}$ are determined from the {\sc mesa} stellar evolution code \citep{paxton11, paxton13, paxton15, paxton18, paxton19} at the zero-age main sequence with solar metallicity, and are $\simeq 0.28 R_{\odot}$ and $\simeq 2.04 R_{\odot}$, respectively.  The points in this figure give the locations where the rate drops to 10\% of $\dot{N}$ for a Schwarzschild SMBH, where $\dot{N}$ is shown by the black, dot-dashed curve, and for the 0.3, 1, and 3 $M_{\odot}$ stars are at $M \simeq 6.3\times 10^{6}M_{\odot}$, $2.3\times 10^{7}M_{\odot}$, and $3.8\times 10^{7}M_{\odot}$, respectively. It is clear that near-maximal spin (dashed curves) increases this mass, but only by a factor of $\sim 1.3$.

We emphasize that the rate given in Equation \eqref{ndotsch} is the rate for a \emph{given galaxy}. The \emph{observed}, volumetric rate is modified by a multitude of factors, including the SMBH mass and spin distributions, the variation of the stellar distribution function among galaxies, differences in the scattering rate of stars into the loss cone, and the detectability of the TDE (which itself depends on many factors; we return to this point in the next subsection). Of course, it is the goal to use TDEs to \emph{infer} these properties of SMBHs and galaxies. We defer a detailed calculation and comparison to observations to a future investigation. However, we note that in general the results shown in Figures \ref{fig:Ntde_spin}, \ref{fig:DeltaN}, and \ref{fig:rates} suggest that the SMBH spin distribution only weakly contributes to the variation in the TDE rate as a function of SMBH mass where the rate itself is substantial (in agreement with \citealt{dorazio19}), and is only able to be probed when the detection rate of TDEs is very accurately constrained as a function purely of SMBH mass (i.e., when confounding effects related to the stellar population, etc., can be ruled out). Therefore, in the short term it seems likely that retrieving information about SMBH spin from TDE observations requires detailed modelling of individual events where, for example, variability may be induced by Lense-Thirring precession of the accretion flow \citep[see, e.g., ][and the processes discussed in \citealt{raj21}]{stone12,franchini16,ivanov18}.

\subsection{The definition of $r_{\rm t}$ and ``observable'' TDEs}
\label{sec:rt}
So far we have employed the canonical definition of the tidal radius, $r_{\rm t} = R_{\star}\left(M/M_{\star}\right)^{1/3}$, to write the TDE fraction (or rate) as a function of SMBH mass and stellar properties, though all of the analysis in Sections \ref{sec:schwarzschild} and \ref{sec:kerr} are agnostic to this definition (modulo Footnote \ref{footnote:1}). The exact distance at which the star is completely destroyed by tides depends on a number of factors (e.g., \citealt{guillochon13, gafton15, mainetti17, gafton19, golightly19b, golightly19a, nixon21}), and it was also recently found numerically that disruptions with $\beta$ \emph{above} a sufficiently large value ($\gtrsim 10$) may result in the reformation of a stellar core \citep{nixon22}. Thus, the ``full disruption radius,'' depends on a large number of variables and may not even constitute a single value for a given star. On the other hand, $r_{\rm t} = 1$ corresponds to the distance at which a substantial fraction of the star's mass is lost, largely independent of the stellar properties, even though it may only be a partial disruption. For example, a solar-like $\gamma = 5/3$ polytrope loses $\sim 10\%$ of its mass to the SMBH at $\beta \simeq 0.68$ ($r_{\rm p} \simeq 1.47 r_{\rm t}$; \citealt{miles20}), while the relativistic simulations of \citet{gafton15} find that this value is closer to $\beta \simeq 0.66$ ($r_{\rm p}\simeq 1.51 r_{\rm t}$) for a $4\times 10^{7}M_{\odot}$ SMBH (see their Figure 3) where the pericenter is highly relativistic. For a solar-like, standard model star, \citet{guillochon13, mainetti17} find that 10\% of the stellar mass accretes onto the SMBH at $\beta \simeq 1.15$ ($r_{\rm p} \simeq 0.87 r_{\rm t}$). Therefore, the canonical definition of $r_{\rm t}$ corresponds to TDEs with substantial mass loss and that are therefore more likely to be observable, irrespective of any nuances related to full vs.~partial disruption.

There have also been analytical arguments by \citet{beloborodov92} (see also \citealt{kesden12, stone19, stone20}) that suggest that the stronger tidal field associated with general relativity modestly increases the tidal radius above the estimate that is derived by equating the stellar self-gravitational force to the Newtonian tidal force, i.e., the fiducial tidal radius. Correspondingly, this analysis suggests a somewhat higher limiting mass above which disruptions can no longer take place (e.g., \citealt{beloborodov92} derive $\sim 10^{8}M_{\odot}$). However, as one approaches the direct capture radius, the relativistic advance of periapsis angle diverges, which implies that \emph{any} SMBH should be able to disrupt a star if its pericenter distance is sufficiently close to the direct capture radius. But, in these scenarios, close to half of the stellar material would be directly captured by the SMBH, half would be unbound from the system, and a small remaining fraction would be able to circularize and accrete. Thus it seems likely that in these situations there would not be much of an accretion flare, and such TDEs would likely be unobservable.

In general, the precise definition of the tidal radius that one uses should incorporate the type of star, the amount of mass lost and the corresponding luminosity of the event (which is currently not well constrained theoretically), and the sensitivity of the detector, and this process carried out for every type of star for a given galaxy and integrated over the stellar population. The information needed to do this is currently not available; in contrast, we argue that the standard definition of the tidal radius gives a value that we expect to correlate strongly with where TDEs are, on average (i.e.~and e.g., across stellar types, with and without general relativistic effects, including or excluding stellar rotation), detectable.

Finally, in our analysis we focused on stars that are disrupted from the region of parameter space where the loss cone is full, meaning that the distribution of the square of the angular momentum of the disrupted stars is effectively uniform. There is another region of parameter space -- the empty regime -- where stars slowly diffuse across the loss cone boundary over many orbital times. Thus, one could argue that our inferred rate suppression is artificially high, as these additional disruptions would enhance the rate near $r_{\rm p} \simeq r_{\rm t}$. However, it is not clear that these would constitute TDEs that yield detectable emission, as it seems likely that such stars would be progressively stripped of their mass over many pericenter passages, leading to underluminous events spread out over long (humanly inaccessible) timescales (as suggested by \citealt{macleod12}); this is especially true at the high SMBH mass where the fallback time of the material becomes $\gtrsim$ years. It is also not possible to substantially reduce the orbital period of such starts through traditional tidal dissipation owing to the extreme mass ratio \citep{cufari22}. Therefore, we expect the rate suppression derived here to be substantial, even with the empty loss cone regime included, though we leave a detailed investigation of the importance of the latter regime to future work.

\section{Summary and Conclusions}
\label{sec:summary}
Under standard assumptions about the nature of the distribution of stars at large distances from the SMBH (Section \ref{sec:boltzmann} and Appendix \ref{sec:appendix}), we analyzed the distribution function of the pericenter distance of tidally disrupted stars, both in the Schwarzschild geometry (Section \ref{sec:schwarzschild}), which can be done completely analytically, and in the Kerr metric (Section \ref{sec:kerr}). Because of the existence of the direct capture radius -- the distance within which the star is unable to escape from the gravitational field of the hole, which in general is distinct from and greater than the horizon distance -- we find that the relativistic distribution function falls off more steeply than would be predicted from a Newtonian analysis for pericenter distances $\lesssim 10\, r_{\rm g}$, where $r_{\rm g} = GM/c^2$ and $M$ the SMBH mass. In particular, for a Schwarzschild SMBH the distribution function equals zero at $4\, r_{\rm g}$ — the direct capture radius for a zero-binding-energy orbit — and is small, but non-zero, for closer distances when the SMBH is rapidly rotating; in the limit that the spin approaches $a = 1$, our analysis demonstrates that the distribution function of the pericenter distance scales as $f_{\rm r_{\rm p}} \propto r_{\rm p}^{4/3}$ or, in terms of the often-defined quantity $\beta = r_{\rm t}/r_{\rm p}$, $f_{\rm \beta} \propto \beta^{-10/3}$. This result can be contrasted with the Newtonian expectation, being that $f_{\rm r_{\rm p}}$ is constant or that $f_{\beta} = \beta^{-2}$.  

As a corollary to this analysis, we derived the total number of TDEs, defined to be those that enter within the tidal radius $r_{\rm t}$ and outside of the direct capture radius, relative to the total number scattered into the loss cone of the SMBH (i.e., all those with pericenter distance within the tidal radius). We find that the total number of TDEs is weakly dependent on the spin of the SMBH, as most clearly shown in Figures \ref{fig:Ntde_spin} and \ref{fig:DeltaN}, and only when the rate of TDEs is very accurately constrained at the high-mass end -- where the intrinsic rate is low but not identically zero if the SMBH is rapidly rotating -- can useful inferences of the SMBH spin be made (see Figure \ref{fig:rates}).

If the stellar population is dominated by low-mass stars, we predict a sharp decline in the rate of TDEs at a SMBH mass that is closer to a value of $\sim 10^{7}M_{\odot}$, even if the SMBH is rapidly rotating. This conclusion disagrees with the inferences of \citet{vanvelzen18}, who from a statistical analysis of 12 SMBH masses inferred from observed TDEs\footnote{Note that \citet{wevers19}, who used a larger sample of TDEs, find that the majority of flares occur at SMBH masses $\lesssim 10^{6}M_{\odot}$.} find a suppression near $10^{8}M_{\odot}$, and also the work of \citet{dorazio19}, who came to a similar conclusion from theoretical grounds. Note, however, that \citet{dorazio19} assumed that the fraction of tidally disrupted stars is 1 if $r_{\rm t}$ exceeds the direct capture radius for a given stellar type, and 0 otherwise, which effectively amounts to setting the direct capture radius to zero if the tidal radius is larger than the direct capture radius (i.e., their solution for $N_{\rm tde}/N$ is a Heaviside step function $H[r_{\rm t}-r_{\rm dc}(a)]$). Given Figure \ref{fig:Ntde_Ndc} and Equations \eqref{Ntde} and \eqref{Ndc}, this approximation clearly and dramatically overestimates the number of TDEs, particularly at the high-mass end. Therefore, from our analysis we expect the rate of TDEs to decline substantially at a mass closer to $\sim 10^{7}M_{\odot}$ rather than $\sim 10^{8}M_{\odot}$. 

Our distribution functions of the pericenter distances of stars scattered into the loss cone of a SMBH, and the total number of tidally disrupted stars derivable therefrom, can serve as direct inputs to calculations of the rates of TDEs. In Section \ref{sec:tderate} we considered the rates per galaxy for individual-star populations, and generalizations to any stellar population are easily derived from Equation \eqref{ndotsch}, given $N_{\rm tde}/N$ from our analysis here. However, as also discussed in Section \ref{sec:tderate}, predictions for the observed rates rely on several assumptions that we have not explicitly made in the work presented here, including (but not limited to) the underlying distribution of SMBH masses and spins, the stellar populations within galactic nuclei, and the precise definition of an observable TDE (which depends on, e.g., the timescale of circularization, the radiative efficiency of accretion, and the cosmological depth of the specific observatory). We leave detailed predictions of observed TDE rates, and their dependencies on these quantities, to future work.

\acknowledgements
We thank the referee for a useful and constructive report. E.R.C.~acknowledges support from the National Science Foundation through grant AST-2006684, and a Ralph E.~Powe Junior Faculty Enhancement Award through the Oakridge Associated Universities. C.J.N.~acknowledges support from the Science and Technology Facilities Council [grant number ST/W000857/1]. 

\appendix
\section{Distribution of specific angular momenta}
\label{sec:appendix}
Any given star is described by its position vector $\mathbf{r}$, its velocity vector $\mathbf{v}$, and the cross product $\mathbf{r}\times\mathbf{v}$. The square of the specific angular momentum can be written in a coordinate-independent way as:

\begin{equation}
J^2 = \left(\mathbf{r}\times\mathbf{v}\right)^2 = \left(\mathbf{r}\cdot\mathbf{r}\right)\left(\mathbf{v}\cdot\mathbf{v}\right)-\left(\mathbf{r}\cdot\mathbf{v}\right)^2 = r^2v^2\left(1-\cos^2\theta\right),
\end{equation}
where here $v$ and $r$ are the magnitude of $\mathbf{v}$ and $\mathbf{r}$ and $\theta$ defines the projection of the velocity vector onto $\mathbf{r}$, i.e., the $r$ component of the velocity is $v_{\rm r} = v\cos\theta$. If the velocity distribution is isotropic, then there is no preferred direction of the velocity, which is ensured if the distribution of $\cos\theta$ is uniform from \{-1, 1\} or equivalently if $f_{\theta}(\theta) = \sin\theta/2$, where $f_{\theta}(\theta)$ is the distribution function of $\theta$. Then the marginalized distribution function of $J^2$ over angle is

\begin{equation}
f_{\rm J^2}(J^2) = \int_{0}^{\pi}\delta\left[J^2-r^2v^2\left(1-\cos^2\theta\right)\right]\frac{1}{2}\sin\theta d\theta,
\end{equation}
and straightforward manipulation of the $\delta$ function yields

\begin{equation}
f_{\rm J^2}(J^2) = \frac{1}{4r^2v^2}\left(1-\frac{J^2}{r^2v^2}\right)^{-1/2}. \label{Jsquaredtot}
\end{equation}
Note that $f_{\rm J^2}(J^2) = 0$ if $J^2 > r^2v^2$. If stars enter the loss cone from $ r \sim 1$ pc with a velocity of $v \sim 100$ km s$^{-1}$, then $J_{\rm lc}^2/(r^2v^2) \sim 10^{-6}$ (setting $J_{\rm lc}^2 = 2GMr_{\rm t}$ as an estimate of the angular momentum necessary to reach the tidal radius), and stars that are scattered into the loss cone have an effectively uniform distribution in $J^2$. 

The projection of the angular momentum onto the spin axis of the hole is given by

\begin{equation}
\ell = \left(\mathbf{r}\times\mathbf{v}\right)\cdot \hat{\mathbf{z}} = \sqrt{J^2}\cos\psi,
\end{equation}
where $\psi$ is the angle between the spin axis of the SMBH and $\mathbf{r}\times\mathbf{v}$. Again, if there is no preferred direction of $\mathbf{r}\times\mathbf{v}$ and the stellar distribution is isotropic, the marginalized distribution of the projection of $\mathbf{r}\times\mathbf{v}$ onto the spin axis should be uniform, and $\cos\psi$ is therefore distributed uniformly and independently of $J^2$; then the joint probability distribution function of $\ell$ and $J^2$ is

\begin{equation}
f(\ell,J^2) = \int \delta\left[\ell-\sqrt{J^2}\cos\psi\right]f_{\rm J^2}(J^2)\frac{1}{2}\sin\psi d\psi.
\end{equation}
As above, we can manipulate the $\delta$-function and geometrically show that this integral evaluates to 

\begin{equation}
f(\ell,J^2) = \frac{1}{\sqrt{J^2}}f_{\rm J^2}(J^2) \label{jointpdf}
\end{equation}
for $\ell^2 < J^2$ and zero for $\ell^2>J^2$. 

\section{Probability formalism in the Kerr metric}
\label{sec:appB}
Here we present the calculations that involve integrals of the joint probability distribution function in the Kerr metric and from which the results in Section \ref{sec:kerr} are derived.

\begin{figure}[htbp] 
   \centering
   \includegraphics[width=0.47\textwidth]{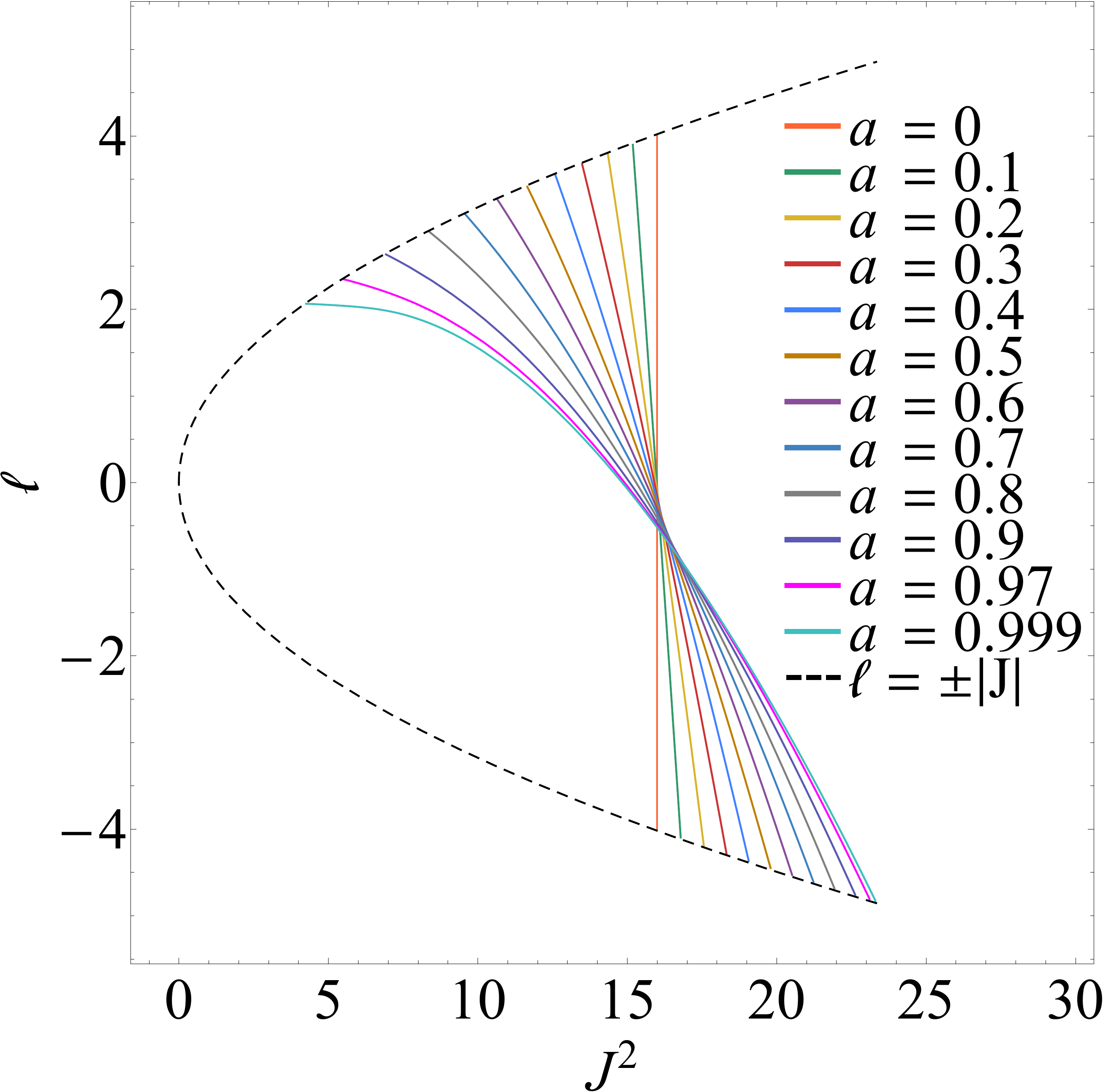} 
    \includegraphics[width=0.49\textwidth]{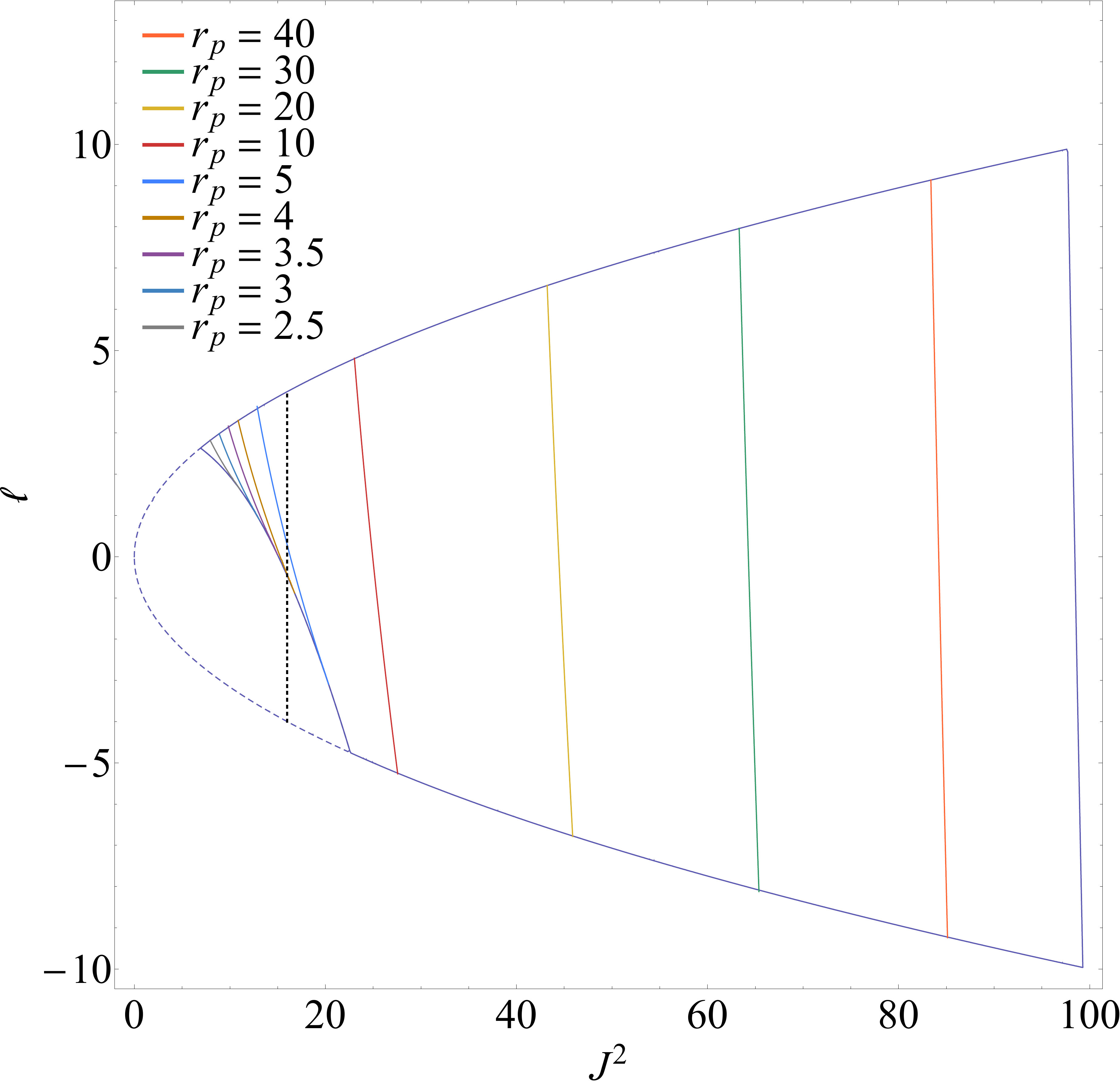} 
   \caption{Left: The direct capture curves for the SMBH spins shown in the legend; for a given $a$, the region of direct capture is bounded by the relevant (colored) curve and the black, dashed curve, which gives $\ell = \pm|J|$ and yields the maximum and minimum value that $\ell$ can have for any $J^2$. When the spin is zero, the curve is a vertical line at $J^2 = 16$, while prograde (positive-$\ell$) orbits can reach smaller $J^2$ without being directly captured when the SMBH is spinning (and retrograde orbits are captured at even larger values of $J^2$). Right: The curves of constant $r_{\rm p}$ in $\ell-J^2$ space, with $r_{\rm p}$ shown in the legend, for a SMBH with $a = 0.9$. For sufficiently small $r_{\rm p}$ the contours intersect the direct capture curve. }
   \label{fig:dccurves}
\end{figure}

For a star to be tidally disrupted, its pericenter distance must be outside the direct capture radius of the SMBH, or within the region of $\ell$-$J^2$ parameter space that is outside the direct capture curve given by Equation \eqref{dccond}. Figure \ref{fig:dccurves} illustrates the direct capture curves for the SMBH spins shown in the legend, such that all points in $\{\ell,J^2\}$ space bounded by the black, dashed curve (which illustrates $\ell = \pm\sqrt{J^2}$ and gives the maximum possible $\ell$ for a given $J^2$) and the colored curve for a given $a$ are directly captured. The solution for $a = 0$ is a vertical line at $J^2 = 16$, i.e., the solution is independent of $\ell$ when the SMBH has no angular momentum because there is no preferred axis, and this agrees with the results of Section \ref{sec:schwarzschild}. As the spin increases, prograde orbits (those with positive $\ell$) can reach smaller $J^2$ and avoid direct capture, while retrograde orbits can be directly captured at $J^2>16$. 

From the left panel of Figure \ref{fig:dccurves} it is obvious that the area of the direct capture region, which is approximately proportional to the number of stars that are directly captured, is not strongly affected by the SMBH spin. There is an effect that is related to the shift of $\ell = 0$ orbits to smaller $J^2$ as $a$ increases, which demonstrates that the area of the direct capture region decreases slightly as $a$ increases. This behavior is due to the presence of $a^2$ terms in Equation \eqref{kofrp}, and a straightforward series expansion of the direct capture curve about $a = 0$ (via Equation \ref{dccond}) shows that the leading-order, spin-dependent term of the area of the direct capture region is $\propto a^2$.  It is straightforward to show with the analysis below that, for a SMBH spin of $a = 0.999$, the relative fraction of directly captured stars drops to $\sim 90.2\%$ (e.g., see Figure \ref{fig:DeltaN}, which shows that the total enhancement in TDEs can be as large as $\sim 10\%$ for maximally spinning SMBHs). 

Observable TDEs must also have a pericenter distance $r_{\rm p}$ less than $r_{\rm t}$, which can be imposed by solving the cubic (from rearranging Equation \ref{kofrp}) and setting $r_{\rm p}(\ell,J^2) \le r_{\rm t}$. The right panel of Figure \ref{fig:dccurves} shows the curves of constant $r_{\rm p}$ in $\ell$-$J^2$ space with $a = 0.9$ and $r_{\rm t} = 47$, with the curves corresponding to the $r_{\rm p}$ given in the legend; the vertical, black, dashed line shows $J^2 = 16$, such that regions of parameter space less than this value are directly captured for Schwarzschild SMBHs. The dashed, purple line shows the direct capture curve, which is identical to the purple curve in Figure \ref{fig:dccurves}. We see that for large $r_{\rm p}$, curves of constant $r_{\rm p}$ coincide roughly with curves of constant $J^2$, the reason being that the spin of the SMBH is only important for highly relativistic encounters. On the other hand, highly relativistic encounters can be achieved for positive $a\ell$ when $J^2 < 16$, while negative $a\ell$ require $J^2 > 16$ to reach the same $r_{\rm p}$ without being directly captured.

With Equation \eqref{jointpdf}, the distribution of pericenter distances satisfies\footnote{Note that when $r_{\rm p}(\ell,J^2)$ is independent of $\ell$, we can integrate over $\ell$ and we recover the same result that we did in the limit of a Schwarzschild SMBH, i.e., we integrate $\ell$ from $-\sqrt{J^2}$ to $+\sqrt{J^2}$ and recover a uniform distribution in $J^2$.}

\begin{equation}
f(r_{\rm p},\ell,J^2) = \delta(r_{\rm p}-r_{\rm p}(\ell,J^2))\frac{f_{\rm J^2}}{\sqrt{J^2}},
\end{equation}
where $r_{\rm p}(\ell,J^2)$ is the solution to Equation \eqref{kofrp}. The cumulative distribution function of $r_{\rm p}$ with observable TDEs is found by integrating $f(\ell,J^2)$ over the $\ell$-$J^2$ space that yields pericenters less than $r_{\rm p}$ and is outside the direct-capture region. From the right panel of Figure \ref{fig:dccurves} we see that there are two global points in this space that are independent of $r_{\rm p}$ that have particular importance, which are the extrema in $\ell$ along the direct capture curve; these are $\ell_{\rm dc, min}$ and $\ell_{\rm dc, max}$, being the minimum (negative) and maximum possible angular momenta along the direct capture curve, and are

\begin{equation}
\ell_{\rm dc, min} = -2\left(1+\sqrt{1+a}\right), \quad \ell_{\rm dc, max} = 2\left(1+\sqrt{1-a}\right), \label{ellminmax}
\end{equation}
and $J^2 = \ell^2$ in both cases, which can be derived from Equation \eqref{dccond} with $\ell = \pm \sqrt{J^2}$. The radii corresponding to these values of $\ell$ and $J^2$ are

\begin{equation}
r_{\rm dc, min} = \left(1+\sqrt{1+a}\right)^2, \quad r_{\rm dc, max} = \left(1+\sqrt{1-a}\right)^2. \label{rmin}
\end{equation}
Note that the radius $r_{\rm dc, max}$ is the smallest possible radius able to be reached by the star without being directly captured, but that this value of the radius corresponds to the maximum value of the specific angular momentum on the direct capture curve, hence the subscript-max. In the maximal spin case with $a = 1$, we have $r_{\rm dc, max} = 1$, which coincides with the horizon. Equations \eqref{ellminmax} and \eqref{rmin} were also obtained by \citet{will12}. 

\begin{figure*}[htbp] 
   \centering
   \includegraphics[width=0.495\textwidth]{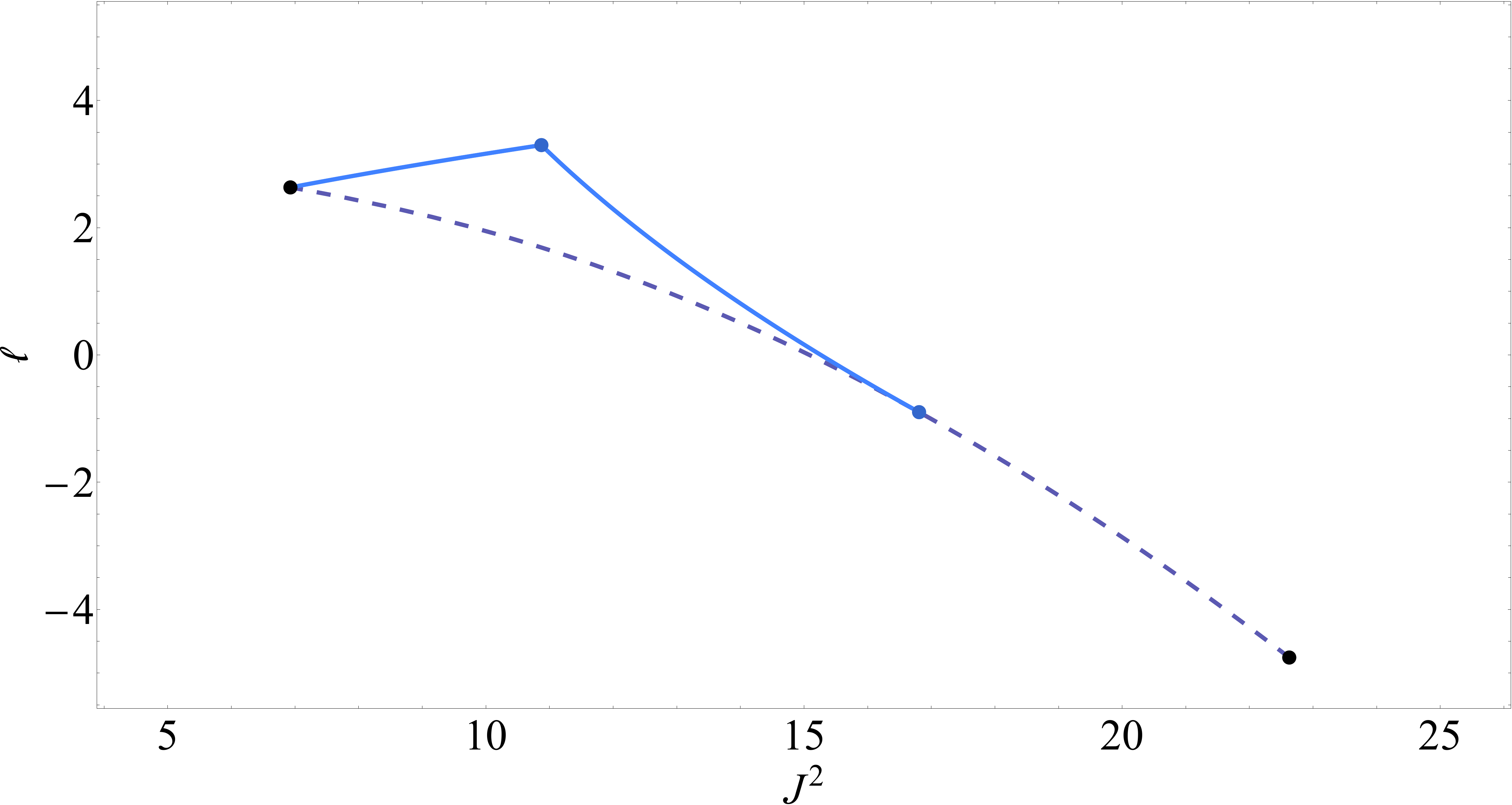} 
   \includegraphics[width=0.495\textwidth]{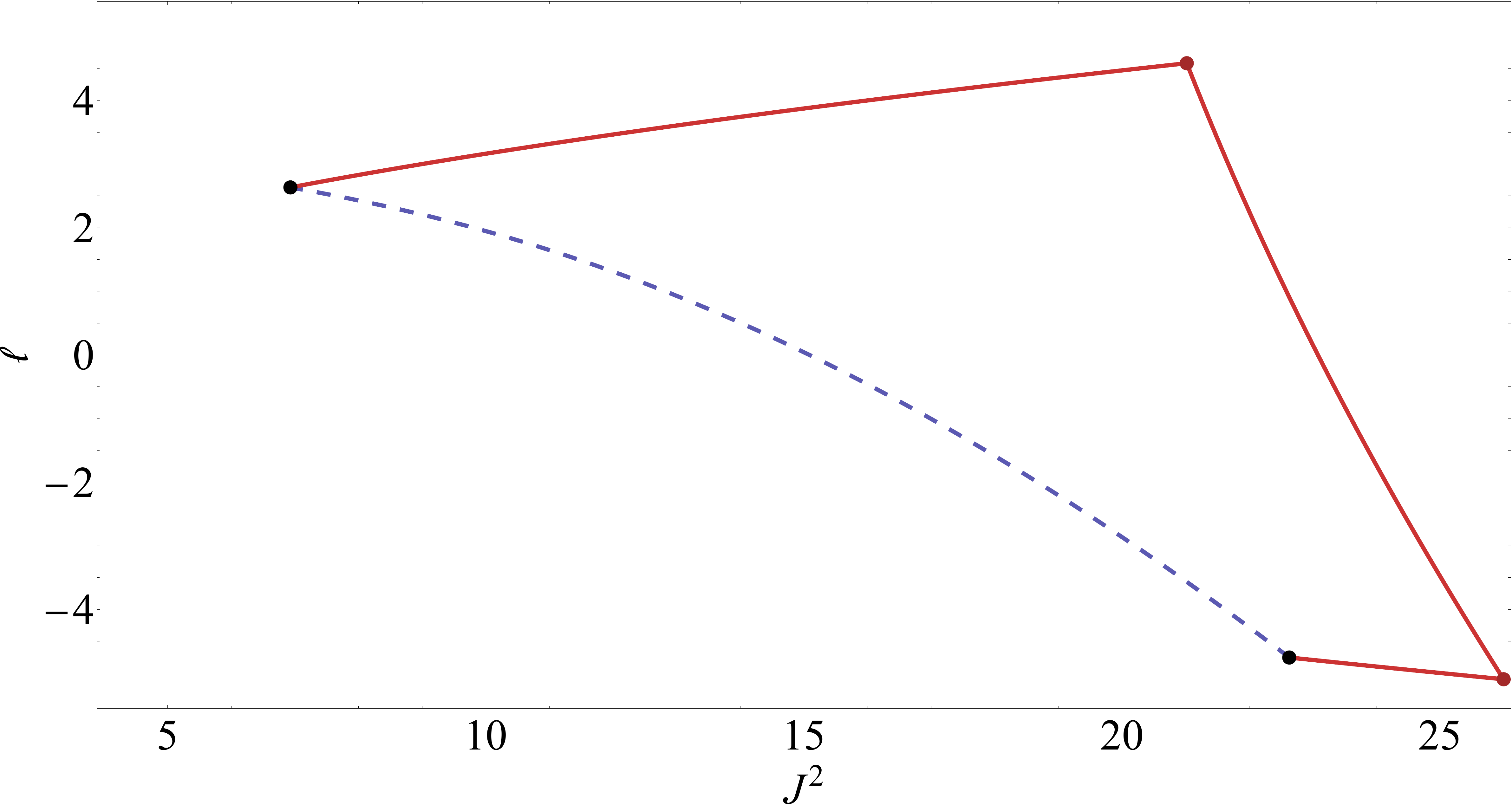} 
   \caption{The regions in $\ell$-$J^2$ space that delimit observable TDEs, i.e., have a pericenter distance less than a given $r_{\rm p}$ but are not directly captured. In the left panel the curve of constant $r_{\rm p} = 4$ is shown by the blue line connecting the two blue points, while the right panel has $r_{\rm p} = 9$ on the curve connecting the two red points. For $r_{\rm p} = 4$, the region of observable TDEs is bounded by the two blue points and the top-most black point (which is the minimum-possible $J^2$ and smallest pericenter; see Equation \ref{rmin}), while that for $r_{\rm p} = 9$ is bounded by the two red points and the two black points.}
   \label{fig:uell}
\end{figure*}

There are two other points that delineate the region of observable TDEs that, unlike Equations \eqref{ellminmax} which depend only on the SMBH spin, also depend on $r_{\rm p}$. If $r_{\rm p} > r_{\rm dc,max}$, these are the minimum and maximum values of $\ell$ with $J^2 = \ell^2$ and a given $r_{\rm p}$; these are, from Equation \eqref{kofrp} with $J^2 = \ell^2$, 

\begin{equation}
\ell_{\rm rp, \pm} = \frac{-2a\pm\sqrt{2r_{\rm p}\left(r_{\rm p}^2-2r_{\rm p}+a^2\right)}}{r_{\rm p}-2}.
\end{equation}
On the other hand, if $r_{\rm p} < r_{\rm dc,min}$, then there will be a minimum-possible $\ell$ that the star can have without being directly captured; this minimum $\ell$ is

\begin{equation}
\ell_{\rm dc, r_{\rm p}} = \frac{a^2+r_{\rm p}^2-2 r_{\rm p}^{3/2}}{a-a\sqrt{r_{\rm p}}},
\end{equation}
and points with $\ell < \ell_{\rm dc, r_{\rm p}}$ are captured. 

Figure \ref{fig:uell} gives two examples to illustrate the regions bounded by these points in angular momentum space; the left panel has $a = 0.9$ and the curve connecting the two blue points corresponds to a constant pericenter distance $r_{\rm p} = 4$. With $a = 0.9$, $r_{\rm dc, max} \simeq 5.657$ (see Equation \ref{rmin}), and hence the curve of constant $r_{\rm p}$ intersects the direct capture curve (shown by the purple, dashed line); there are thus three points that bound the region of observable TDEs with $r_{\rm p} < 4$. The right panel has $a = 0.9$ and the curve connecting the red points has $r_{\rm p} = 9$, and in this case the curve of constant $r_{\rm p}$ does not intersect the direct-capture curve; hence there are four points that delineate the region in which observable TDEs occur with $r_{\rm p} < 9$. In each of these panels the top-most black point shows $\ell_{\rm dc, max} \simeq 2.632$, while the bottom-most black point corresponds to $\ell_{\rm dc, min} \simeq -4.757$ (and in both cases $J^2 = \ell^2$; see Equation \ref{ellminmax}). The cumulative distribution function of observable TDEs for a given $r_{\rm p}$ (i.e., all TDEs with pericenter distances less than $r_{\rm p}$ but outside the direct capture region) is the integral of the joint distribution function, given by Equation \eqref{jointpdf}, over these regions.

\bibliographystyle{aasjournal}

\end{document}